\documentclass{aa}   

\usepackage{graphicx}
\usepackage{upgreek}
\usepackage{txfonts}

\newcommand{\CII}{C\,{\sc ii}}
\newcommand{\HII}{H\,{\sc ii}}

\begin{document} 

\title{Molecular globules in the Veil bubble of Orion\thanks{Based on IRAM 30m telescope observations. IRAM is supported by INSU/CNRS (France), MPG (Germany), and IGN (Spain).}}

\subtitle{IRAM\,30\,m $^{12}$CO, $^{13}$CO, and C$^{18}$O\,(2-1) expanded maps of Orion~A}

 \titlerunning{Globules in the Veil bubble of Orion} 
\authorrunning{Goicoechea et al.} 
                                      
 \author{J.\,R.\,Goicoechea\inst{1}
          \and
        C.\,H.\,M.\,Pabst\inst{2} 
          \and
        S.\,Kabanovic\inst{3} 
          \and
        M.\,G.\,Santa-Maria\inst{1} 
          \and
        N.\,Marcelino\inst{1} 
          \and
        A.\,G.\,G.\,M.\,Tielens\inst{2} 
          \and
        A.\,Hacar\inst{2} 
          \and
        O.\,Bern\'e\inst{4}
          \and
        C.\,Buchbender\inst{3}
          \and
        S.\,Cuadrado\inst{1}
          \and
        R.\,Higgins\inst{3}
            \and
        C.\,Kramer\inst{5}
          \and
        J.\,Stutzki\inst{3}
          \and
        S.\,Suri\inst{6}
          \and
        D.\,Teyssier\inst{7}
          \and
        M.\,Wolfire\inst{8}
          }

 \institute{Instituto de F\'{\i}sica Fundamental
     (CSIC). Calle Serrano 121-123, 28006, Madrid, Spain.
              \email{javier.r.goicoechea@csic.es}
         \and
Leiden Observatory, Leiden University, Leiden, The Netherlands.          
         \and
I. Physikalisches Institut der Universit\"at zu K\"oln, Cologne, Germany.             
         \and  
IRAP, Universit\'e de Toulouse, CNRS, CNES, Universit\'e Paul Sabatier, Toulouse, France.         
        \and
Institut de Radioastronomie Millim\'etrique (IRAM), Grenoble, France.        
        \and
Max Planck Institute for Astronomy, Heidelberg, Germany.        
        \and
Telespazio Vega UK Ltd. for ESA/ESAC, Madrid, Spain        
        \and
University of Maryland, Astronomy Department, College Park, MD, USA.
            }
   
   \date{Received 8 January 2020 / accepted 27 April  2020}


\abstract{Strong winds and ultraviolet (UV) radiation from  O-type  stars disrupt and ionize their molecular core birthplaces, sweeping up \mbox{material} into parsec-size shells.  Owing to dissociation by starlight, the thinnest  shells are expected to host low molecular abundances and therefore little star formation.  
Here, we expand previous maps made with observations using  the IRAM\,30m telescope 
\mbox{(at 11$''$\,$\simeq$\,4,500\,AU resolution)} and present
 \mbox{square-degree} \mbox{$^{12}$CO and $^{13}$CO ($J$\,=\,2-1)} maps of the
wind-driven  \mbox{``Veil bubble''} that surrounds the \mbox{Trapezium} cluster and its natal Orion molecular core (OMC). 
Although widespread and extended CO  emission is largely absent from the Veil, 
we show that several CO ``globules'' exist that  are blueshifted in velocity with respect to OMC and are  embedded in the [\CII]\,158\,$\upmu$m-bright shell that confines the bubble. This includes the first detection  of quiescent CO at negative local standard of rest (LSR) velocities in Orion. 
Given the  harsh UV irradiation conditions in this translucent material, the detection of CO globules is surprising.  These globules  are small 
(\mbox{$R_{\rm g}$\,$=$\,7,100\,AU),} not massive (\mbox{$M_{\rm g}$\,$=$\,0.3\,$M_\odot$}), and are moderately dense: \mbox{$n_{\rm H}$\,$=$\,4$\cdot$10$^{4}$\,cm$^{-3}$} (median values). They are 
confined by the external  pressure of the  shell, $P_{\rm ext}/k$\,$\gtrsim$\,10$^7$\,cm$^{-3}$\,K, and are likely magnetically supported. They are  either transient objects formed by instabilities  or have detached from pre-existing molecular structures, sculpted by the passing shock associated with the expanding shell and by  UV radiation from the Trapezium. Some represent the first stages in the formation of small pillars, others of isolated small globules. Although their  masses ($M_{\rm g}$\,$<$\,$M_{\rm Jeans}$) do not suggest they will form stars, one  globule matches the position of a known young stellar object.  
The lack of extended CO  in the  ``Veil shell'' demonstrates that feedback from massive stars expels, agitates, and reprocesses most of the disrupted molecular cloud gas, thereby
limiting the star-formation rate in the region.  
The presence of molecular globules  is a result of this feedback.
}

\keywords{galaxies: ISM – H II regions – ISM: bubbles – ISM: clouds
--- ISM: individual (Orion)}

   \maketitle
%

\section{Introduction}

Massive stars dominate the injection of UV radiation into the \mbox{interstellar} medium (ISM) and of mechanical energy through stellar winds and supernova explosions. 
The energy and momentum injected by photoionization, radiation pressure, and stellar winds from young O-type stars ionize and disrupt their natal molecular cloud cores, creating 
\HII~regions and blowing parsec-size bubbles enclosed by shells of denser swept-up material \citep[e.g.,][]{Weaver77,Churchwell06,Deharveng10}. 
These feedback processes may locally regulate the formation of new stars, and globally drive the evolution of the ISM in galaxies as a whole
\mbox{\citep[e.g.,][]{Krumholz14,Rahner17,Haid18}}.

\begin{figure*}[ht]
\centering   
\includegraphics[scale=0.175, angle=0]{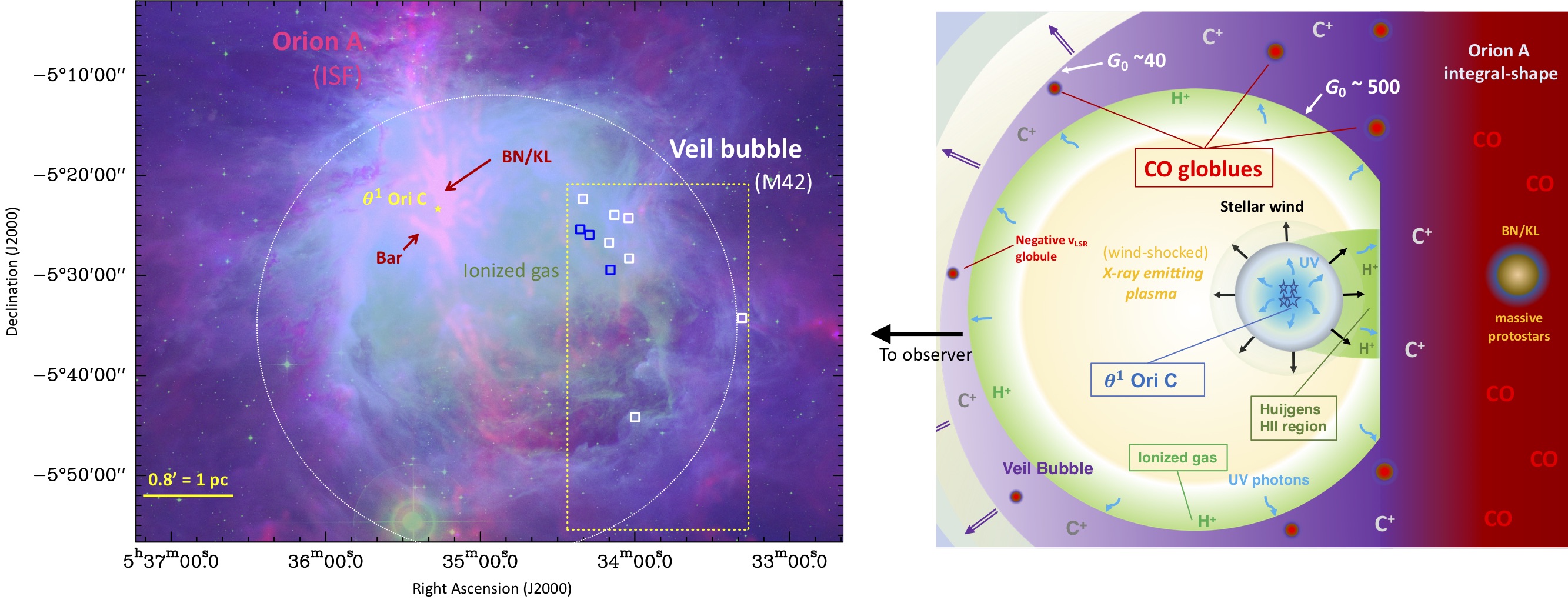}
\caption{\textit{Left}: Extended Orion Nebula (M42), part  of the integral-shape filament
in Orion\,A  (reddish colors), and the  Veil bubble (delineated by a dotted  circle) 
 filled with ionized gas (greenish). 
Red:~SPIRE\,500\,$\upmu$m image (cold dust from the background molecular cloud). Blue:~PACS\,70\,$\upmu$m image (warm dust).
Green: H$\alpha$ image adapted from ESO's  Second Digitized Sky Survey   \citep[DSS2, see][]{Pabst20}.
 The yellow star  corresponds to the position of \mbox{$\theta^1$ Ori C} in the \mbox{Trapezium} cluster. Here we focus on the \mbox{17.5$'$\,$\times$\,34.5$'$} area enclosed by the dotted box. The blue and white squares mark the position of \mbox{CO globules} detected in the expanding shell that confines the bubble (blue squares for the \mbox{negative-v$_{\rm LSR}$} globules). \textit{Right}:~Sketch of the region, not at scale, adapted from \cite{Pabst19}.
 }\label{fig:finder}
\end{figure*} 

The iconic Extended Orion Nebula (M42) is photoionized by UV photons emitted mainly from the most massive star in the Trapezium cluster, \mbox{$\theta^1$ Ori C} \citep[type O7V and \mbox{$Q_{\rm Ly}$\,$\simeq$\,6$\cdot$10$^{48}$\,photons\,s$^{-1}$}; e.g.,][]{Odell01,Simon06,Karl18}. In addition, the characteristic bubble-shape ($\sim$4\,pc in diameter) and overall dynamics of M42 seem ultimately driven by the strong wind emanating from \mbox{$\theta^1$ Ori C} \citep{Gudel08,Pabst19}.  
The foreground  material that  surrounds the Trapezium  and its natal \mbox{molecular core-1} \citep[\mbox{OMC-1}, located behind the cluster, e.g.,][]{Genzel89,Bally08} are generically  known as the ``Veil'' 
\citep[][]{Odell01,vanderWerf_2013,Troland16,Abel19}. 
The \mbox{``Veil bubble''} is filled with an \mbox{X-ray-emitting} \mbox{(wind-shocked)} million-degree plasma \citep{Gudel08}. As delineated by the   H$\alpha$ emission, the inside of this bubble is also an \HII~region photo-ionized by UV radiation from 
\mbox{$\theta^1$ Ori C}. In the far side, the bubble  is confined by dense molecular gas at the surface of \mbox{the OMC}  \citep[][]{Rodriguez98,Goico19} and, in the near side, by an expanding half-shell of warm gas ($T_{\rm k}$$\simeq$100\,K) and dust  (see sketch in Fig.~\ref{fig:finder}). The swept-up material  in the shell is very bright in the \mbox{$^2P_{3/2}$\,-\,$^2P_{1/2}$} fine-structure emission of C$^+$ (the well-known [\CII]\,158\,$\upmu$m line). 
In \citet{Pabst19} we  obtained square-degree velocity-resolved images  of the
[\CII]\,158\,$\upmu$m emission with SOFIA, and showed  that the mechanical energy from the stellar wind from \mbox{$\theta^1$ Ori C} \citep[terminal velocity of \mbox{$\sim$2,500\,km\,s$^{-1}$}; ][]{Stahl96} is effectively converted into kinetic energy of the shell.  This stellar wind  causes more disruption of \mbox{OMC-1} (well before any supernova explosion) than photo-ionization or photo-evaporation \citep{Pabst19}.

Although not the most numerous  or massive star cluster in the Milky Way, the proximity of the Orion nebula, the Trapezium stars, and the molecular core \mbox{OMC-1} (the closest core that hosts ongoing
massive-star formation) enables us to study  star formation and stellar feedback  
in great spatial  detail \citep[like in our previous works, here we adopt $\sim$414\,pc; e.g.,][]{Menten07}. 

As in other thin shells around high-mass stars, the intense  UV radiation in M42 suggests very low molecular abundances in the ``Veil shell''. Indeed, previous observations 
of the line of sight toward the Trapezium stars imply small columns of material  in the Veil \citep{Odell01},  \mbox{1-2\,mag} of visual extinction ($A_{\rm V}$), and also  low molecular gas fractions 
\citep[$x$(H$_2$)/$x$(H)\,$<$\,10$^{-4}$; where $x$ is the abundance with respect to  H nuclei,][]{Abel06}.

 CO, the second most abundant molecule in the ISM, had not been detected toward the Veil before. As the emission from cold H$_2$ is not directly observable either  \citep[e.g.,][]{Bolatto13}, the lack of detectable CO emission poses uncertain constraints on the measurable mass of molecular material that escapes detection in wide-field CO radio surveys \citep{Grenier05,Planck11}. This extended ``CO-dark'' molecular gas (when the CO column density, $N$(CO), is too low to be detected) may represent 30\%~of the molecular gas mass in the Milky Way
\citep{Grenier05,Wolfire10}. This fraction can be much higher in the ISM of low-metallicity galaxies \citep{Madden97} characterized by a higher penetration of stellar UV radiation.  In this context, the Orion’s Veil is an interesting nearby template to study the origin and properties of the vast neutral halos that surround many star-forming regions. 

This paper is organized as follows. In Sect.~2 we  describe the new  \mbox{CO ($J$\,=\,2-1)} mapping  observations. In Sect.~3 we present the main observational result of this work, namely the detection of globules embedded in the shell that confines the Veil bubble. In Sect.~4 we analyze the environment and the properties of these globules, and in Sect.~5 we discuss their  origin and evolution.
  
\begin{figure*}[t]
\centering   
\includegraphics[scale=0.176, angle=0]{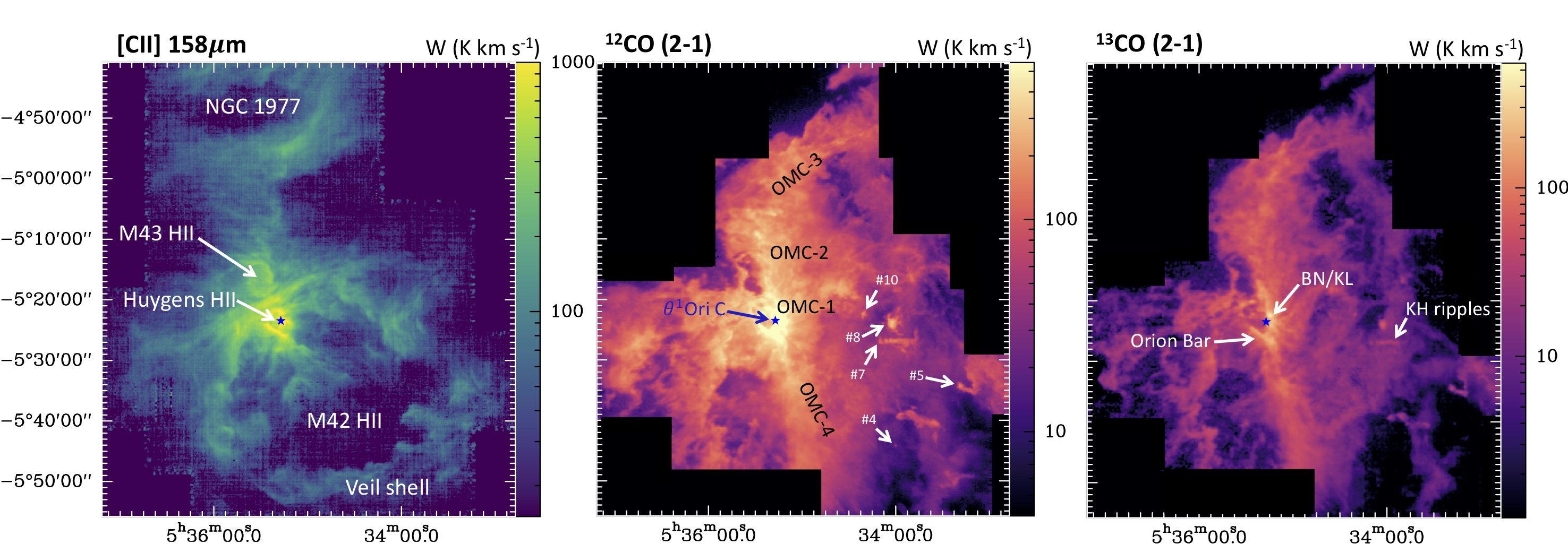}
\caption{[\CII]\,158\,$\upmu$m, $^{12}$CO~(2-1), and $^{13}$CO~(2-1) integrated line intensity maps of the central square-degree region of Orion~A. Some of the main structures and components discussed in this work are labelled.  The [\CII]\,158\,$\upmu$m map
was observed by SOFIA/UPGREAT at an agular resolution of 16$''$ \citep[][]{Pabst19}. The $^{12}$CO and $^{13}$CO maps, observed with the IRAM\,30m telescope, have a resolution of 11$''$.}\label{fig:complete-CO-maps}
\end{figure*}

\section{Observations and data reduction}\label{sec-observations}

We obtained new $^{12}$CO, $^{13}$CO, and C$^{18}$O~\mbox{($J$\,=\,2-1)} 
\mbox{fully sampled} maps of Orion~A using the IRAM 30m telescope (Pico Veleta, Spain). The bright central region \mbox{($1^{\circ}\times0.8^{\circ}$)} around \mbox{OMC-1} was originally mapped in 2008 \citep{Berne14} with the  multi-beam receiver HERA at  0.4~km\,s$^{-1}$ resolution  \citep{Schuster04}.
In order to cover the larger area (1.2~square-degree) mapped by us in the [\CII]\,158\,$\upmu$m line with SOFIA/GREAT  at comparable angular resolution (Legacy Program led by \mbox{A.~G.~G.~M.~Tielens})
we started to expand the CO maps using  EMIR \citep{Carter12} and  FFTS backends
at the 30m telescope. These new observations of fainter regions in Orion~A were carried out in October 2018, March 2019, November 2019, and February 2020, so far employing $\sim$100\,h of telescope time, and are part of the Large Program ``Dynamic and Radiative \mbox{Feedback} of Massive Stars'' \mbox{(PI: J.~R.~Goicoechea).} 

The $^{12}$CO $J$=2-1 (230.5\,GHz), $^{13}$CO $J$=2-1 (220.4 GHz), and  
C$^{18}$O $J$=2-1 (219.5\,GHz) lines were simultaneously mapped with  EMIR, providing an instantaneous bandwidth of 16\,GHz per polarization, in combination with FFTS backends at 200\,kHz  resolution 
($\sim$0.25\,km\,s$^{-1}$). The half power beam width (HPBW) at 230.5\,GHz
is 10.7$''$.   The observing strategy consisted in mapping boxes of $\sim$534$''$$\times$534$''$ size using the \mbox{on-the-fly} (OTF) technique. Each box was mapped through 16 rectangular tiles of 8 OTF scans each, followed by a calibration measurement. In total, we employed about 12 min  per tile. The scanning velocity was $\sim$9$''$\,s$^{-1}$ and the dump time was $\sim$0.4\,s. This combination provided three dumps per beam along the scanning direction. The separation between two successive raster lines was 4.2$''$ (roughly beam/2.5) with the scanning direction reversed after each raster line (i.e., zigzag scanning mode). To avoid edge-effects between consecutive boxes, we overlapped the observation of each box edge by slightly more than one beam.  In order to decrease stripping, each box was mapped twice, scanning in two orthogonal directions. We used a common reference position that we call REF, at an offset $(-60', -30')$ from the map center.  We obtained deep spectra of the REF position using the frequency switching mode. These spectra do not show  CO emission at the rms level of the map. The REF position was observed for 10 s after each raster line 
(taking $\sim$60\,s) following the pattern REF--OTF--OTF--REF. Pointing was checked every 2\,h and focus every 4\,h or after sunset. We started every day with
a pointed observation of the Orion Bar to cross-check the intensity of the CO lines. We estimate an absolute intensity error of about 15\%. 

The new data were first calibrated in the $T_{\rm A}^{*}$ scale, and were then corrected for atmospheric absorption and spillover losses  using the chopper-wheel method \citep{Penzias73}. 
Most of these observations were made under very good winter  conditions (less than \mbox{2 mm} of atmospheric precipitable water vapor). The  receiver and system temperatures
were typically $\sim$100\,K and \mbox{$\sim$200-300\,K}, respectively.
For emission sources of  brightness temperature \mbox{$T_{\rm b}$(v)} at a given velocity channel, the main-beam temperature is the most appropriate intensity scale (i.e., \mbox{$T_{\rm mb}({\rm v}) \simeq T_{\rm b}({\rm v})$}) when the emission source fills the main beam of the telescope.
Assuming a disk-like emission source  of uniform $T_{\rm b}$   and angular size varying from $\sim$25$''$ to 90\,$''$, the ratio
$T_{\rm mb}$\,/\,$T_{\rm b}$ at 230\,GHz goes from  1.0 to 1.2  at the IRAM\,30m telescope \citep[whereas $T_{\rm A}^{*}$\,/\,$T_{\rm b}$ goes from 0.4 to 0.5; see Online material in][]{Teyssier02}. Molecular clouds have complicated emission structures, spatially and in velocity, meaning that the $T_{\rm mb}$ scale is widely used as a good compromise when the emission sources are smaller than the
very wide antenna error beams \citep[for the IRAM\,30m telescope, see][]{Greve98}. Here we converted the intensity scale from $T_{\rm A}^{*}$ to   
 $T_{\rm mb}$ ($= T_{\rm A}^{*} \cdot F_{\rm eff}/B_{\rm eff}$) using the main-beam efficiency $B_{\rm eff}$ and forward efficiency  $F_{\rm eff}$ appropriate for each frequency ($B_{\rm eff}$\,=\,0.59 and $F_{\rm eff}$\,=\,0.92 at 230\,GHz).

  \begin{figure*}[th]
\centering   
\includegraphics[scale=0.22, angle=0]{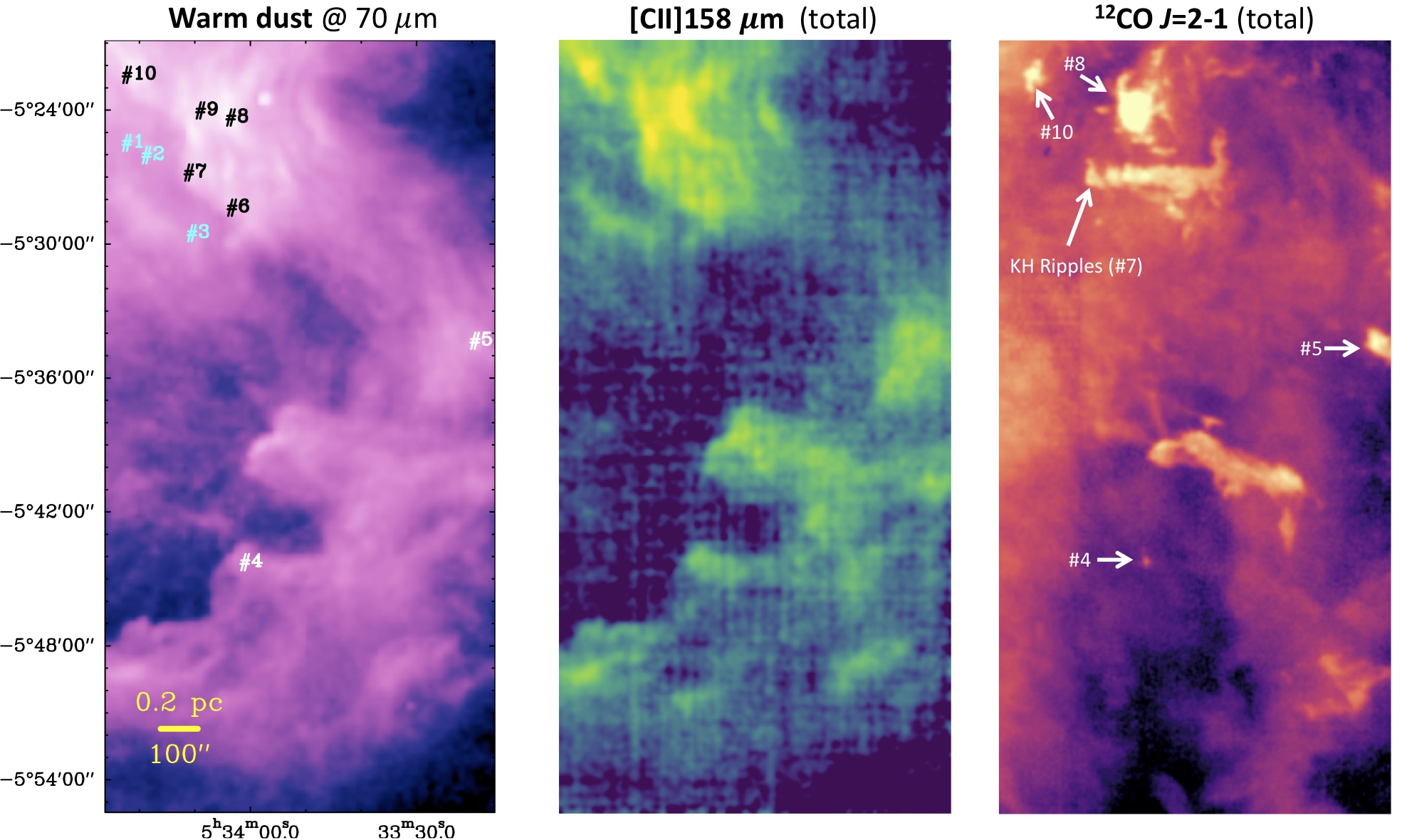}
\caption{Zoom on the southwest region of the Veil shell. The first two images
 show the emission from FUV-heated warm dust and from C$^+$. 
 The rightmost image displays a rather different morphology, mostly dominated by extended CO from the molecular cloud behind the shell.  \textit{Left panel:} PACS\,70\,$\upmu$m emission at 6$''$ resolution and positions of the detected CO globules. \textit{Middle panel:} 
SOFIA  [\CII]\,158\,$\upmu$m intensity map integrated in the 
\mbox{v$_{\rm LSR} = [-7,+20]$\,km\,s$^{-1}$} range \citep{Pabst19}.  
\textit{Right panel:} IRAM\,30\,m   $^{12}$CO~(2-1) intensity map in the v$_{\rm LSR} = [-7,+20]$\,km\,s$^{-1}$ range. We identified blueshifted CO globules (some of them are labeled) with velocity centroids in the v$_{\rm LSR}$ range $[-7,+6]$\,km\,s$^{-1}$ (see Fig.~\ref{fig:globules}). 
}\label{fig:rgb}
\end{figure*}

Data reduction was carried out with the GILDAS software\footnote{http://www.iram.fr/IRAMFR/GILDAS}.     
         A polynomic baseline of order 1 or 2 was subtracted avoiding  velocities with molecular emission. Finally, the spectra were gridded into a data cube through a convolution with a Gaussian kernel of  approximately one-third the telescope HPBW.         
The typical (1$\sigma$) rms noise level  achieved in the map is 0.25\,K per 0.25\,km\,s$^{-1}$ velocity channel.  This is typically a factor of more than three deeper than the rms of the large $^{12}$CO~(1-0) map obtained by merging CARMA interferometric and NRO 45m telescope observations at \mbox{10$''$$\times$8$''$} angular resolution \citep{Kong18}. 
Although these latter authors detect the stronger positive-velocity globules and other bright structures (their Fig.~8), the sensitivity in our maps allowed us to investigate faint and diffuse \mbox{CO~(2-1)} emission structures and compact globules at local standard of rest (LSR) velocities 
significantly blueshifted from those of the OMC. 

In order to compare with our SOFIA [\CII]\,158\,$\upmu$m map,
 we merged the older CO HERA observations with the expanded EMIR maps. Thanks to the improved new  software, \mbox{MRTCAL,} at the 30\,m telescope  with calibration on a finer  frequency grid, line calibration has slightly improved since 2017.
In addition, telescope efficiencies have slightly changed as well. 
Hence, we took the new EMIR data as the reference for the CO line intensities. To do that, we re-observed a few common areas and  produced scatter plots of the CO line integrated intensities \mbox{(HERA vs. EMIR maps)}. The derived linear slopes  deviate by $<15\,\%$, and we used this correction factor to scale up the HERA data. Figure~\ref{fig:complete-CO-maps} shows the (current) extent of the $^{12}$CO and \mbox{$^{13}$CO (2-1)} merged maps, whereas Fig.~\ref{fig:rgb} zooms into the  area of interest for this work.         

In order to properly compare the \mbox{$^{12}$CO (2-1)}, \mbox{$^{13}$CO (2-1)}, and [\CII]\,158\,$\upmu$m line profiles at the same angular and spectral resolutions, 
we created cubes convolved, with a Gaussian kernel, to  uniform resolutions of 16$''$ and 0.4\,km\,s$^{-1}$, respectively. The  convolved and smoothed CO maps
were used in the line profile analysis (see Sect.~\ref{sec-analysis} and Fig.~\ref{fig:spectra} for the spectra) as well as in the
 position--velocity diagrams 
(Fig.~\ref{fig:pv_4} and \ref{fig:pv_1} to \ref{fig:pv_10}). 
The typical  rms noise in the smoothed $^{12}$CO (2-1) map is 0.16~K per 0.4\,km\,s$^{-1}$ channel. We used Gaussian fits to extract the line profile parameters of the blueshifted CO globules: spectral components peaking at LSR velocities lower than those of \mbox{OMC} 
\citep[i.e.,~\mbox{v$_{\rm LSR}$$<$$+$(7-10)\,km\,s$^{-1}$}; ][]{Bally87,Berne14,Kong18}. 
Line fit parameters   are tabulated in Tables~\ref{table:line-peaks} and \ref{table:line-fits} of the Appendix.
When appropriate, offsets in arcsec are given with respect to star  \mbox{$\theta^1$ Ori C},  at \mbox{$\alpha(2000)=$05h35m16.46s} and  \mbox{$\delta(2000)=-05^o23'22.8''$}.

\section{Results}

\begin{figure*}[t]
\centering   
\includegraphics[scale=0.185, angle=0]{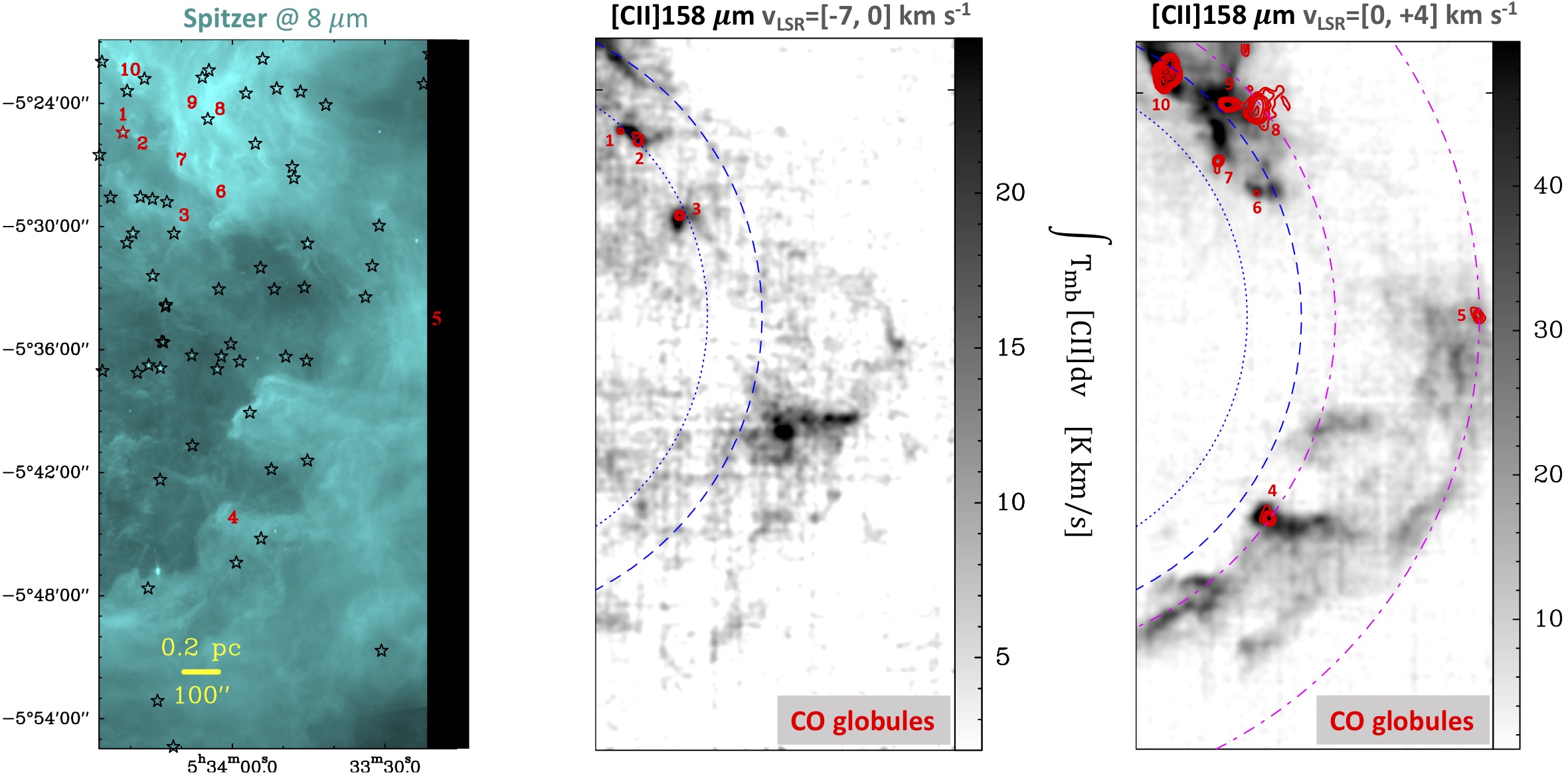}
\caption{CO globules in the  shell that encloses the Veil bubble. \textit{Left panel:} Spitzer's 8\,$\upmu$m image (2$''$ resolution). Numbers mark the position of the CO globules (shown in the middle and right panels). Stars show the position of YSOs detected in the field \citep{Megeath12,Megeath16}. Globule \#1 matches the position of YSO \#1728 (red star).  \textit{Middle panel:} [\CII]\,158\,$\upmu$m emission at  v$_{\rm LSR}=-7$ to 0\,km\,s$^{-1}$ (background grey image) and $^{12}$CO  globules \#1, \#2 and \#3 in the same velocity range (emission in red contours). \mbox{\textit{Right panel:}}  Same as the middle panel but for the range v$_{\rm LSR}$\,=\,0 to $+$4\,km\,s$^{-1}$. The arcs approximately represent the projection of concentric expanding rings in the shell (blue for rings closer to us).}\label{fig:globules}
\end{figure*}

Figure~\ref{fig:complete-CO-maps} shows a square-degree area of the
Orion~A molecular complex. At visible wavelengths, the region is dominated by M42, the extended Orion nebula, ionized by the
strong UV radiation field from  \mbox{$\theta^1$ Ori C} \citep[e.g.,][]{Odell01}. While in this region the [\CII]\,158\,$\upmu$m emission mostly traces UV-illuminated gas around
 \HII~regions, the majority of the \mbox{$^{12}$CO~(2-1)} and \mbox{$^{13}$CO~(2-1)} integrated line emission  arises from the molecular cloud behind (see sketch Fig.~\ref{fig:finder}).
The properties of the main star-forming cores (\mbox{OMC-1}, \mbox{OMC-2}, \mbox{OMC-3},
and \mbox{OMC-4}) have been extensively discussed in previous CO maps of the region
\citep[e.g.,][]{Bally87,Shimajiri11,Buckle12,Berne14,Kong18}.
Here we focus on a \mbox{$\sim$17.5$'$\,$\times$\,34.5$'$} \mbox{($\sim$2\,pc\,$\times$\,4\,pc)} region of the bubble (zoomed in Fig.~\ref{fig:rgb}) southwest from the Trapezium. The Veil shell is very conspicuous at 70\,$\upmu$m  (warm grains), in the [\CII]\,158\,$\upmu$m line \citep{Pabst19}, and in the extended  8\,$\upmu$m emission produced by polycyclic aromatic hydrocarbons 
 (PAHs, Fig.~\ref{fig:globules}\,left).  However, the $^{12}$CO integrated line intensity map shows a rather different morphology, which is dominated by emission structures in the background dense molecular cloud. This  region of ongoing star formation, part of  Orion’s integral-shape filament, dominates the CO emission and peaks at local standard of rest velocities (v$_{\rm LSR}$) around 
\mbox{$+$(7-10)\,km\,s$^{-1}$} \citep[e.g.,][]{Bally87,Shimajiri11,Buckle12,Berne14,Kong18}.
 At these positive LSR velocities, high-resolution \mbox{(sub-km\,s$^{-1}$)} spectra of the [\CII]\,158\,$\upmu$m line display bright emission from the UV-irradiated surface of the dense molecular cloud
\citep{Boreiko96,Ossenkopf13,Goico15,Goico19,Cuadrado19}.

\begin{figure*}[t]
\centering   
\includegraphics[scale=0.62, angle=0]{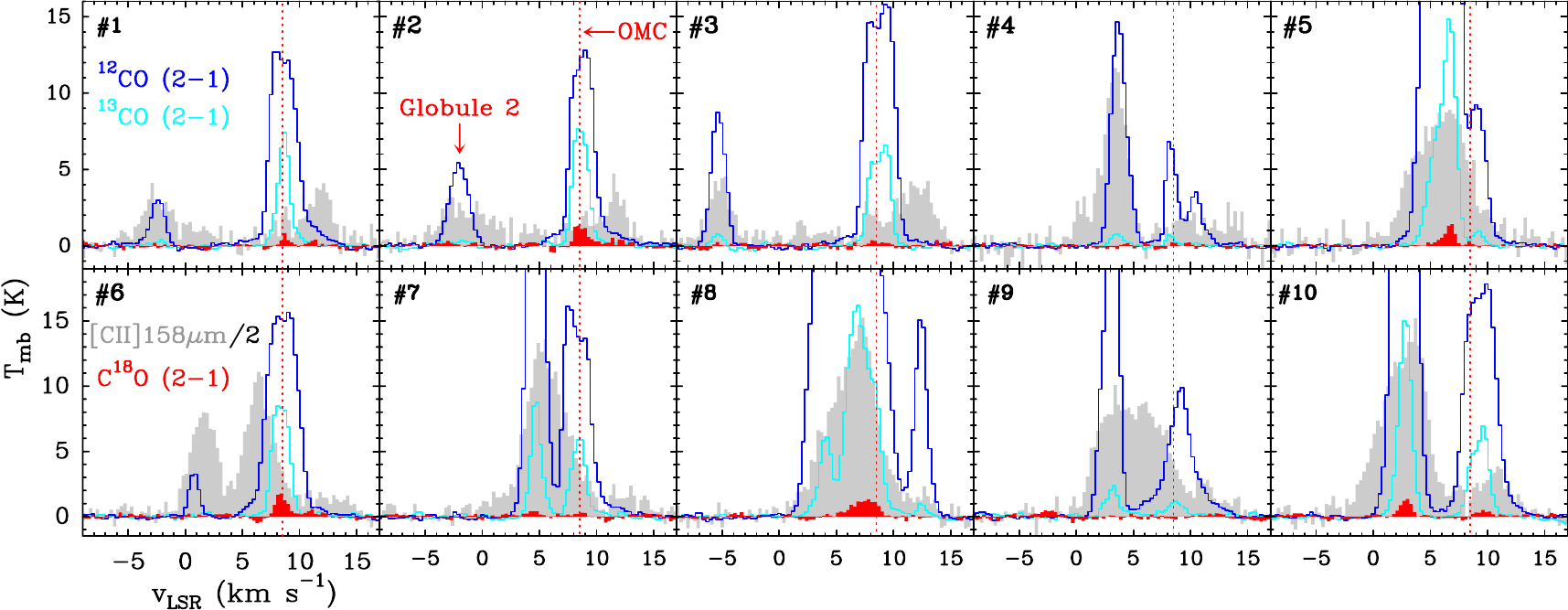}
\caption{Velocity-resolved spectra toward the CO globules in Orion's Veil. Each panel shows the [\CII]\,158\,$\upmu$m line (colored in gray), $^{12}$CO\,$J$=2-1 (dark blue), 
$^{13}$CO $J$=2-1 (cyan), and C$^{18}$O $J$=2-1 (red) lines toward  the emission peak of each globule (labeled by symbol \# as in \mbox{Figs.~\ref{fig:rgb} and \ref{fig:globules}}). The x-axis represents the LSR velocity in 
km\,s$^{-1}$. The vertical red dotted line marks the approximate velocity of the emission produced by OMC and associated star-forming molecular cloud located behind the Veil. All panels show emission features at  velocities blueshifted from OMC. Globules \#1, \#2, and \#3 represent the first detection of quiescent CO emission structures at negative LSR velocities in Orion.}\label{fig:spectra}
\end{figure*}

 In addition, the [\CII]\,158\,$\upmu$m spectra reveal fainter  
  components that are blueshifted from \mbox{OMC-1} velocities and reach \mbox{negative} 
LSR velocities (see the spectra in Fig.~\ref{fig:spectra}). \citet{Goico15} already found that $\sim$15\,\% of the [\CII]\,158\,$\upmu$m luminosity toward the central regions
of \mbox{OMC-1} does not arise from its \mbox{UV-illuminated} surface.
 This C$^+$  emission mainly comes from the foreground  half-shell that surrounds \mbox{OMC-1} and expands \mbox{(toward us)} at 13\,km\,s$^{-1}$ \citep{Pabst19, Pabst20}.  The relatively narrow [\CII]\,158\,$\upmu$m line profiles ($\Delta$v\,$\approx$\,4\,km\,s$^{-1}$) in the  shell demonstrate that the gas  is largely neutral. That is, dominated by H, H$_2$, and C$^+$ (with an ionization potential of 11.3\,eV). Indeed, hydrogen recombination lines from fully ionized \HII~regions display much broader profiles \citep[$\Delta$v\,$>$15\,km\,s$^{-1}$ for $T_e$\,$>$\,5,000\,K;][]{Churchwell78}. 
\mbox{However, carbon recombination} lines from the surface and edges of \mbox{OMC-1}, detected at radio \citep{Natta94,Salas19} and  \mbox{millimeter} wavelengths \citep{Cuadrado19},  show \mbox{narrow} profiles
\mbox{$\Delta$v\,=\,2.5-5\,km\,s$^{-1}$}. These are typical of the neutral 
\mbox{photodissociation region} (PDR) that separates the hot \HII~gas from the cold molecular gas. 
The 8\,$\upmu$m emission from UV-pumped PAHs also arises from PDR gas
\mbox{\citep[e.g.,][]{Hollenbach97}}.
Hence, the good correlation between the  C$^+$ and  PAH emission from the shell \citep{Pabst19}, together with the narrow [\CII]\,158\,$\upmu$m line-widths, supports the conclusion that most of the  C$^+$ emission in the shell originates from neutral PDR gas rather than in the ionized gas.

\subsection{Detection of blue-shifted CO globules in the Veil }

The dotted rectangular box in Fig.~\ref{fig:finder} shows the specific area investigated in this work (expanded in Fig.~\ref{fig:rgb}). Despite the 0.75\,K\,km\,s$^{-1}$ (3$\sigma$) sensitivity level of our  $^{12}$CO~(2-1) map, equivalent to    \mbox{$N$($^{12}$CO)\,$\gtrsim$\,(5-15)$\cdot$10$^{14}$\,cm$^{-2}$},  we do not detect widespread and extended CO emission from the shell (i.e.,  blueshifted  from the \mbox{OMC}). For the typical extinction \mbox{($A_{\rm V}$\,$\sim$\,1-2\,mag)} and plausible gas densities 
($n_{\rm H}$ of several 10$^3$\,cm$^{-3}$) in this  material \citep{Odell01,Abel16}, UV photodissociation must severely restrict the formation of abundant CO 
\mbox{\citep[][]{Dishoeck88}}. Nevertheless, the lack of detectable extended CO emission does not directly imply that the whole shell is 100\% atomic throughout and not molecular.

Indeed, Fig.~\ref{fig:globules}  
shows the presence of globules and other CO emission structures blueshifted  from \mbox{the OMC}. This includes the first detection of quiescent molecular gas at negative v$_{\rm LSR}$  (\#1, \#2, and \#3) and well separated from the main star-forming cores in the integral-shape filament  \mbox{\citep[][]{Johnstone99,Kirk17}}. We extracted the angular
sizes of these globules using the native \mbox{CO} maps at 11$''$ resolution. They range from 18$''$ ($\sim$7,500\,AU) to 80$''$ ($\sim$0.16\,pc), with their fainter emission contour typically above 5$\sigma$. 

\begin{figure*}[h]
\centering   
\includegraphics[scale=0.213, angle=0]{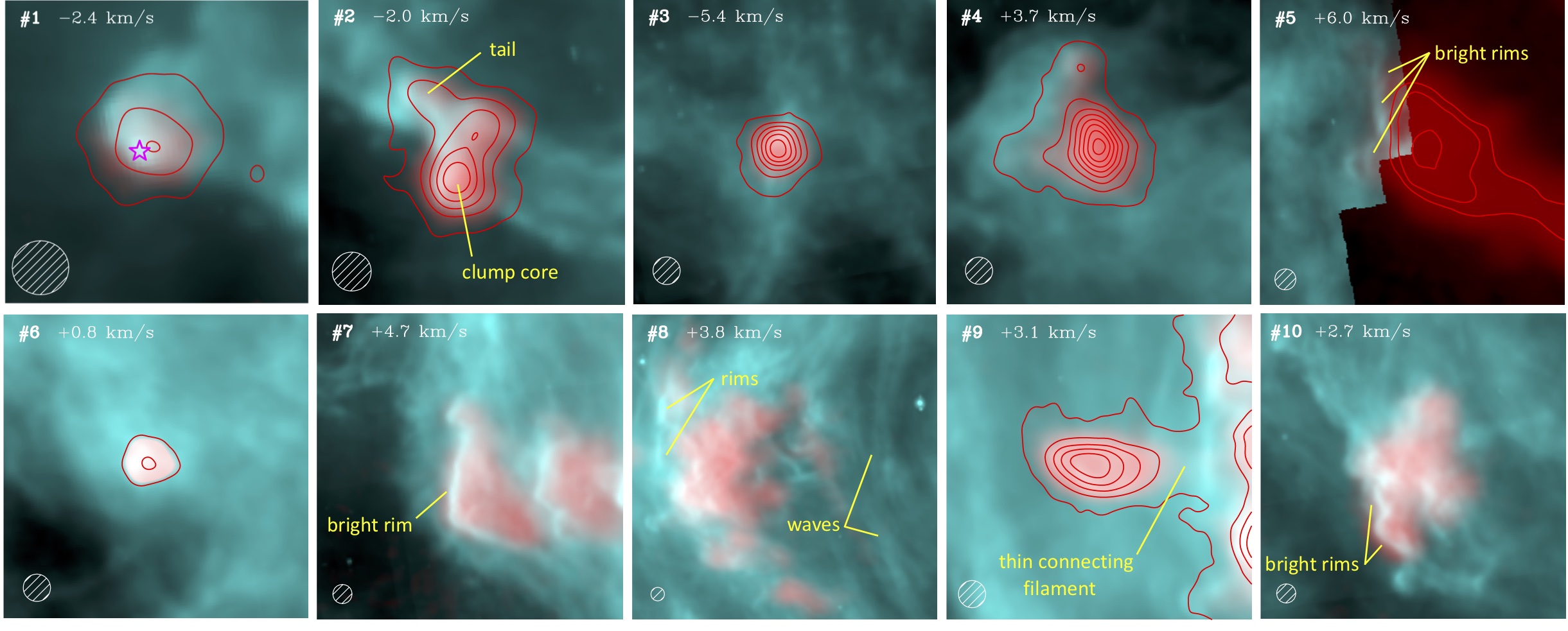}
\caption{Gallery of blueshifted CO globules and emission structures detected toward Orion's Veil bubble. The reddish color  is the  \mbox{$^{12}$CO~(2-1)} emission
integrated over the appropriate emission velocity range of each globule (spectra shown in Fig.~\ref{fig:spectra}). The 
bluish color is the 8\,$\upmu$m emission (extended PAH emission) imaged with Spitzer/IRAC
at 2$''$ resolution.
 Each panel indicates the LSR velocity centroid of the CO line-profiles  and a white-lined circle  with  the 11$''$ ($\sim$4,500\,AU) beam size of the CO
observations. The images of the smaller CO globules display
\mbox{$^{12}$CO~(2-1)} intensity contours (in red) starting with the 5$\sigma$ rms level (except for the brighter globules \#5 and \#9). 
In order to exhibit  the bright rims (\mbox{FUV-illuminated} edges) of the larger and brighter CO globules, their images do not display CO emission contours (except for globule \#5, where the contours help to locate the head of a more elongated structure). Globule \#1 matches the position of YSO \#1728 \citep[magenta star,][]{Megeath12}. Figure~\ref{fig:globule-images-Cplus} shows the same gallery
but displaying the velocity-resolved [\CII]\,158\,$\upmu$m emission in bluish.}\label{fig:globule-images}  
\end{figure*}

Figure~\ref{fig:globule-images} shows a gallery with zooms on the 
globules and other blueshifted structures in the \mbox{CO~(2-1)} \mbox{(reddish)} and 8\,$\upmu$m band  (bluish) emission.
Figure~\ref{fig:globule-images-Cplus} of the Appendix  shows the same gallery
but displaying the [\CII]\,158\,$\upmu$m emission (bluish) integrated in exactly the same velocity range as the CO emission from each globule 
\mbox{(i.e., C$^+$ that is strictly connected in velocity with CO)}.
The smallest globules (\#1, \#2, \#3, and \#6) 
resemble the kind of tiny clouds originally seen in visible plates against the nebular
emission from  \HII~regions (ionized by radiation from nearby OB stars) and referred to as ``globules''\footnote{Interestingly, no globule associated with the Orion nebula was found in the original work of \citet{Bok47}.} \citep{Bok47,Minkowski49,Thackeray50}, ``cometary globules'', or \mbox{``tear drops''} \citep{Herbig74}, and more recently   ``cusps'' \mbox{\citep{DeMarco06}}  or \mbox{``globulettes''} when their sizes are smaller than about 10,000\,AU \citep{Gahm07}.
Hence, in this work we use the term \mbox{``globule''} in a generic sense to signify 
small and over-dense molecular gas blobs. \mbox{Globules \#1}, \#3, and \#6  show
spherical morphologies but they are  surrounded by extended
 8\,$\upmu$m (and [\CII]\,158\,$\upmu$m) emitting structures. \mbox{Globules \#2} and \#4 are also roundish, but show indications of diffuse tails. These globules have CO velocity centroids that are significantly blueshifted from OMC and are isolated.

In addition, previous CO maps of OMC   reveal more \mbox{extended} and peculiar structures \citep{Shimajiri11,Berne14,Kong18}. Some of our blueshifted but \mbox{positive-v$_{\rm LSR}$} globules belong to these  structures. They are located close to the surface of the  dense molecular cloud
(but are likely of a different nature) or at the limb-brightened edge of the shell.
 A very remarkable structure of this kind is the Kelvin–Helmholtz (KH) \mbox{``ripples''} or \mbox{``periodic undulations''} studied by \citet{Berne10} and indicated in our 
Figs.~2 and 3. \mbox{Globule \#7} is the blueshifted head of the KH~ripple, whereas
 \mbox{\#5} in the westernmost part of the map is the tip of a more prominent and bright-rimmed structure, a \mbox{pillar} or \mbox{"elephant trunk"} that points toward the Trapezium.
The extended structures associated with \#5, \#8 and \#10 also show multiple \mbox{far-UV} (FUV; $E$\,$<$\,13.6\,eV) illuminated edges  revealed by their bright 8$\,$$\upmu$m rims delineating the CO emission. These globules must be facing strong FUV fluxes.
 \mbox{Globule  \#8} has a different morphology. It 
is part of a more extended region that is apparently connected to the OMC, and is
characterized by wavelike structures and  8$\,$$\upmu$m rims pointing toward the Trapezium. The CO and [\CII]\,158\,$\upmu$m emission follows an arched morphology  also roughly pointing toward the Trapezium. Finally, globule \#9 appears to be
detached from \#8, still showing a thin connecting filament.

\begin{figure*}[ht]
\centering   
\includegraphics[scale=0.45, angle=0]{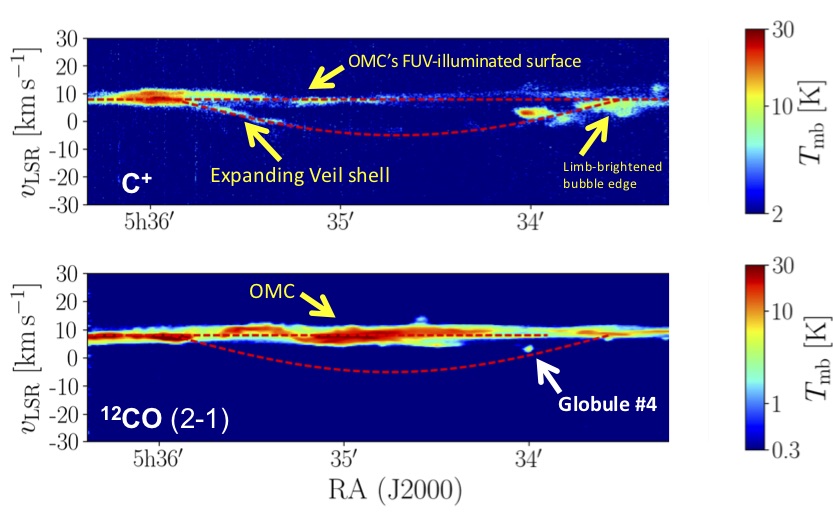}
\caption{Position--velocity diagram of the [\CII]\,158\,$\upmu$m (upper panel) and \mbox{CO (2-1)} (lower panel) emission along an  eastwest cut across the bubble at the declination of globule \#4 (-$5^o44'11.55''$).
We created this diagram by averaging spectra over a cut of 45$''$  wide in declination.
 A~model of a half-shell expanding at 13\,km\,s$^{-1}$ is shown as a curved red dashed line \citep[from][]{Pabst19}. The bright emission at \mbox{v$_{\rm LSR}$\,=\,+(7-10)\,km\,s$^{-1}$} (horizontal red dashed line) corresponds to 
 the background dense molecular cloud and the integral-shape filament.}\label{fig:pv_4}
\end{figure*}

The similar velocity centroid of the CO and [\CII]\,158\,$\upmu$m lines 
toward each globule
 \mbox{(Table~\ref{table:line-fits})}, as well as the spatial coincidence with velocity-coherent C$^+$ extended emission structures indicate that several globules are embedded in the shell  (see position--velocity diagrams in Fig.~\ref{fig:pv_4} and Figs.~\ref{fig:pv_1} to \ref{fig:pv_10}). 
  Owing to the shell expansion, the \mbox{negative-v$_{\rm LSR}$} globules should be located in the near side of the shell (closer to us) and  not at the surface of \mbox{the OMC}.   These globules resemble starless cores embedded in a C$^+$-bright envelope 
(for \mbox{globule \#4}  with the morphology of the [\CII]\,158\,$\upmu$m emission akin to a cometary globule, Fig.~\ref{fig:globules}). Despite their
relatively faint $^{12}$CO emission levels, we also detect $^{13}$CO toward most of them. 
However, with the exception of globules \#5 and \#10,  we do not detect blueshifted C$^{18}$O emission. The lack of  \mbox{C$^{18}$O\,(2-1)} emission is consistent with low extinction depths in the range of A$_{\rm V}$\,$\lesssim$\,3\,mag \citep[][]{Frerking82,Cernicharo87,Pety17}.

 To conclude this general presentation and analysis of the region, in Fig.~\ref{fig:pacs70} we show the approximate map of $G_0$, the stellar FUV flux in the line of sight toward the bubble \citep[$G_0$\,$\simeq$\,1.7 is the mean radiation field in the solar neighborhood in Habing  units;][]{Habing68}.
These FUV photons heat grains, ionize C atoms, and dissociate CO molecules. Figure~\ref{fig:pacs70} shows that the innermost regions of the shell are directly exposed to FUV radiation from the Trapezium stars at $G_0$ levels of several hundred. The FUV flux reaching the outer portions of the shell is more attenuated, down to \mbox{$G_0$\,$\sim$\,40} (see Sect.~\ref{sec-G0} for details).
 
\subsection{Observed line parameters of the CO globules}

The \mbox{$^{12}$CO (2-1)} line profiles  are relatively narrow, with a median value of \mbox{$\Delta$v($^{12}$CO)\,=\,1.5\,km\,s$^{-1}$}. 
These profiles are remarkably Gaussian (except for \#5 and \#8 that display blended components) and do not show the kind of line asymmetries or wings expected in collapsing, or expanding, or outgassing globules.
The velocity centroid of both  $^{12}$CO and $^{13}$CO lines  coincides  for each globule (both species arise from the same gas component). However, the \mbox{$^{13}$CO (2-1)} lines are
narrower, that is, \mbox{$\Delta$v($^{13}$CO)\,=\,1.2\,km\,s$^{-1}$} (median), indicating that the  \mbox{$^{12}$CO (2-1)} lines are opacity-broadened \citep{Phillips79}.
         The [\CII]\,158\,$\upmu$m lines toward the globules are significantly broader, that is,  \mbox{$\Delta$v(C$^{+}$)\,=\,3.2\,km\,s$^{-1}$} (median), and show more intricate line profiles. Assuming \mbox{$T_{\rm k}$=100 K} for the C$^+$ gas \citep{Pabst19}, $T_{\rm k}$=10\,K for $^{13}$CO, and separating the thermal broadening ($\Delta\sigma_{\rm th}$(C$^+$)\,=\,0.26\,km\,s$^{-1}$ and $\Delta\sigma_{\rm th}$($^{13}$CO)\,=\,0.05\,km\,s$^{-1}$), the differences in observed line widths  
 imply that the [\CII]\,158\,$\upmu$m emission arises from a more turbulent gas that surrounds the $^{13}$CO cores, with  \mbox{$\sigma_{\rm turb}$(C$^+$)/$\sigma_{\rm turb}$($^{13}$CO)}\,$\simeq$\,\mbox{1.3/0.5} 
 ($\Delta$v$_{\rm obs}$\,=\,2.35$\sigma_{\rm obs}$ and
$\sigma_{\rm obs}^{2}$\,=\,$\sigma_{\rm turb}^{2}$+$\sigma_{\rm th}^{2}$). These  dispersions imply transonic to supersonic motions (see Table~\ref{table:globule-masses} for each globule).

\begin{table*}[h]  
\caption{Globule parameters estimated from line observations adopting a reasonable range of $^{12}$CO abundances (single-slab approximation).\label{table:LTE-globules}} 
\centering  
\begin{tabular}{rccccccc@{\vrule height 8pt depth 5pt width 0pt}}
\hline\hline
      &  Coordinates                     & $R_{\rm g}$  & $T_{\rm ex}$(CO)$^a$ & $N$(CO)$^b$                     &   $x$(CO)             & $n_{\rm H}$ $^c$ &   $A_{\rm V}$$^d$ \\        
globule &  (Ra. Dec.)                        &     (AU)      & (K)                  & (cm$^{-2}$)                       &  (10$^{-5}$ adopted) & (10$^4$ cm$^{-3}$) & (mag)           \\    
\hline                                                                                                                                                                            
 \#1  & 5h34m21.21s -$5^o25'24.23''$ &  3,700 & 7.6$\pm$0.6  & (3.8$\pm$0.5)$\cdot$10$^{16}$  &  1.5$-$0.5 & 4.8$\pm$2.8  & 1.9$\pm$1.1 \\   
 \#2  & 5h34m17.65s -$5^o25'57.52''$ &  7,000 & 10.0$\pm$0.9 & (same as \#1)                           &  1.5$-$0.5 & 2.5$\pm$1.5  & 1.9$\pm$1.1 \\   
 \#3  & 5h34m09.55s -$5^o29'29.40''$ &  5,800 & 13.8$\pm$1.4 & (3.8$\pm$0.2)$\cdot$10$^{16}$  &  1.5$-$0.5 & 3.0$\pm$1.6  & 1.8$\pm$1.0 \\   
 \#4  & 5h33m59.86s -$5^o44'11.55''$ &  8,300 & 19.8$\pm$2.2 & (3.8$\pm$0.1)$\cdot$10$^{16}$  &  1.5$-$0.5 & 2.1$\pm$1.1  & 1.8$\pm$0.9 \\  
 \#5  & 5h33m18.44s -$5^o34'16.29''$ &  7,500 & 46.2$\pm$6.0 & (1.8$\pm$0.2)$\cdot$10$^{18}$  & 15.0$-$5.0 & 10.9$\pm$6.1 & 8.5$\pm$4.8 \\  
 \#6  & 5h34m02.27s -$5^o28'18.58''$ &  3,100 & 7.9$\pm$0.6  & (2.8$\pm$0.3)$\cdot$10$^{16}$  &  1.5$-$0.5 & 4.3$\pm$2.5  & 1.4$\pm$0.8 \\  
 \#7  & 5h34m10.00s -$5^o26'44.24''$ &  6,200 & 46.4$\pm$6.1 & (5.8$\pm$0.5)$\cdot$10$^{17}$  & 15.0$-$5.0 & 4.3$\pm$2.4  & 2.8$\pm$1.6 \\   
 \#8  & 5h34m02.46s -$5^o24'17.04''$ & 16,600 & 34.4$\pm$4.4 & (3.0$\pm$0.3)$\cdot$10$^{17}$  & 15.0$-$5.0 & 0.9$\pm$0.5  & 1.5$\pm$0.8 \\  
 \#9  & 5h34m07.99s -$5^o23'57.64''$ &  7,200 & 31.8$\pm$4.0 & (1.4$\pm$0.1)$\cdot$10$^{17}$  & 15.0$-$5.0 & 0.9$\pm$0.5  & 0.7$\pm$0.4 \\   
\#10  & 5h34m20.14s -$5^o22'21.93''$ & 14,500 & 58.0$\pm$7.9 & (1.5$\pm$0.2)$\cdot$10$^{18}$  & 15.0$-$5.0 & 4.7$\pm$2.7  & 7.6$\pm$4.6 \\ 
 \hline
 &                              \hspace{3cm}   Median   & 7,100  & 25.8               & 8.6$\cdot$10$^{16}$   &         8.3$-$2.8    &  3.6 & 1.8    \\
 \hline
\end{tabular}
\tablefoot{$^a$Assuming optically thick $^{12}$CO emission and taking into account absolute intensity calibration errors and 1$\sigma$ errors
discussed in Sec.~\ref{sec-observations}. 
$^b$From optically thin $^{13}$CO emission.
$^c$For $n_{\rm H}$\,=\,$N_{\rm H}$/2$R_{\rm g}$ with $N_{\rm H}$\,=\,$N$(CO)/$x$(CO)$_{\rm adopted}$.
$^d$Defined as $A_{\rm V}$\,=\,3.5$\cdot$10$^{-22}$\,$N$($^{12}$CO)/$x$($^{12}$CO)$_{\rm adopted}$.}
\end{table*}

\section{Analysis}\label{sec-analysis}

Wide-field  near-IR  and visible  images of  \HII~regions 
allow the detection of globules of neutral gas and dust in their surroundings \mbox{\citep[e.g.,][]{DeMarco06,Gahm07}}. Detected in silhouette or as  bright cusps, photometric images reveal the  morphology and projected position of these compact objects. Velocity-resolved  C$^+$ and CO spectroscopic images (at \mbox{sub-km\,s$^{-1}$} resolution)  help to find these globules in velocity space, and also  provide a means of quantifying their physical conditions, bulk masses, and gas kinematics.

A few studies of this kind have previously focused on small CO globules seen in planetary nebulae around evolved stars such as the Helix Nebula \citep[e.g.,][]{Huggins92,Huggins02} and on CO globules seen toward  \HII~regions in more distant high-mass star-forming regions
 such as the Rosette Nebula \citep{Schneps80,Gonzalez94,Dent09,Gahm13} located at a distance of 1.6~kpc. 
In this section, we analyze the [\CII]\,158\,$\upmu$m, $^{12}$CO, and $^{13}$CO  emission from the globules detected in Orion at high spatial resolution.

\subsection{Single-slab analysis}

With the detection of a single rotational line, and given the possible  small-scale structure and potential gradients in the physical conditions
of these globules, the derivation of their gas density, temperature, and mass 
is not trivial.
 In~this~section, we assume that the CO level populations  are characterized by a single excitation temperature (i.e., single-slab analysis).
Given the low critical densities for collisional excitation of the observed CO lines \mbox{($n_{\rm cr\,,2-1}$\,$\approx$\,10$^4$\,cm$^{-3}$)}, we implicitly assume that the $^{12}$CO levels are close to thermalization 
 \mbox{($T_{\rm ex}$\,$\simeq$\,$T_{\rm k}$)} and that $^{13}$CO is a good tracer of the total molecular column density. This is obviously a first-order approach
 (e.g., at low densities the \mbox{higher-$J$} lines will be subthermally excited) but it is highly complementary to the mere photometric detection of these globules.  
 In Sect.~\ref{sec:PDR-mods} we perform a more detailed depth-dependent  modeling.

The observed $W$($^{12}$CO)/$W$($^{13}$CO) integrated line intensity ratio  toward all globules is always lower than~25 (with $W$=$\int{T_{\rm mb}\,d{\rm v}}$ in K\,km\,s$^{-1}$; see Fig.~\ref{fig:ratios} in the Appendix).   This is lower than the $^{12}$C/$^{13}$C isotopic ratio ($R_{\rm 12C13}$) in OMC-1 \citep[$R_{\rm 12C13}$=67$\pm$3;][]{Langer90} and implies that: $(i)$ the $^{12}$CO emission is optically thick,  $(ii)$ the $^{13}$CO abundance is enhanced over the expected isotopic ratio, so-called chemical fractionation \citep[][]{Langer84}, or $(iii)$ both. For optically thick lines, the \mbox{$^{12}$CO (2-1)} excitation temperature $T_{\rm ex}$ is a good lower limit on $T_{\rm k}$, and we can directly  extract this from the line peak temperature, $T_{\rm P,12}$, of each globule  (see Table~\ref{table:line-peaks}):
\begin{equation} 
T_{\rm ex}=\frac{h\nu/k}{{\rm ln\,}\left(1 + \frac{h\nu/k}{T_{\rm P,12}+ J(T_{\rm bg})}\right)}, 
\end{equation} 
 where $J(T)=(h\nu/k)\,/\,(e^{h\nu/kT} - 1)$   is the equivalent brightness temperature of a black body at $T$, and $T_{\rm bg}$ is the cosmic background temperature 2.7 K. 
We note that we spatially resolve these globules, and so no beam-filling factor correction is needed. 
Taking into account that \mbox{$T_{\rm ex}$($^{13}$CO 2-1)\,=\,$f$$\cdot$$T_{\rm ex}$($^{12}$CO 2-1)}, where \mbox{$f$\,$\leq$\,1}, 
we can determine the opacity of the \mbox{$^{13}$CO (2-1)} line,  
\mbox{$\tau_{13}$\,=\,$-$ln\,(1$-$$\frac {T_{\rm P,13}}{J(T_{\rm ex})-J(T_{\rm bg})})$},
from observations, and also the $^{13}$CO column density: 
\begin{equation}
N(^{13}{\rm CO}) = N_{{\rm thin}}(^{13}{\rm CO})\frac {\tau_{13}}  {1-e^{-\tau_{13}}},
\end{equation}
 where $N_{\rm thin}$  is the $^{13}$CO column density (in cm$^{-2}$) in the 
 \mbox{$\tau_{13}\rightarrow0$}  limit (see eq.~3). 
 The factor $f$ reflects the possible different excitation temperatures of \mbox{$^{12}$CO (2-1)} and \mbox{$^{13}$CO (2-1)} lines due to line-trapping effects as the \mbox{$^{12}$CO (2-1)} line opacity increases. The parameter $f$ tends to 1 in collisionally excited optically thin gas. We carried out nonlocal thermodynamic equilibrium (NLTE) calculations that show that line-trapping reduces $f$ for optically thick $^{12}$CO emission (roughly above $N$($^{12}$CO) of a few 10$^{16}\,$cm$^{-2}$) and at low $n$(H$_2$) densities (typically lower than  $\sim$10$^4$\,cm$^{-3}$, the critical density of the \mbox{$J$=2-1} transition). For the expected gas densities and $N$(CO) in these globules, we find  $f\simeq\,$0.9 and this is the factor  we use here.
       We calculate $N$($^{13}$CO) assuming a Boltzmann distribution of the  level populations at a uniform $T_{\rm ex}$.  
          In this case,
\begin{equation}          
          N_{\rm thin}=8\pi (\frac{\nu}{c})^3\, \frac{Q(T_{\rm ex})}{g_2 A_{21}}
          \frac{e^{E_2/kT_{\rm ex}}}{e^{h\nu/kT_{\rm ex}}-1} 
          \frac{W}{J(T_{\rm ex})-J(T_{\rm bg})},\end{equation}\label{eq:Nthin}where $W$ is the $^{13}$CO\,(2-1) line integrated intensity for each globule (values tabulated in Table~\ref{table:line-fits}). 

 We derive the $^{12}$CO column density as \mbox{$N$($^{12}$CO)=$R_{\rm 12C13}\cdot N(^{13}$CO)}, and the  column density of gas across each CO globule core,  \mbox{$N_{\rm H}$\,=\,$N$(H)+2$N$(H$_2$)}, defined as 
     \mbox{$N_{\rm H}$\,=\,$N$($^{12}$CO)/$x$($^{12}$CO)}.
Supported by our more detailed  photochemical modeling (following section), we estimate the  extinction ($A_{\rm V}$) through each globule and their mass ($M_{\rm g}$) adopting a plausible  range of (uniform) CO abundances. We use  \mbox{$x$($^{12}$CO)\,=\,(0.5-1.5)$\cdot$10$^{-5}$} when $N$($^{12}$CO)\,$<$\,5$\cdot$10$^{16}$\,cm$^{-2}$ (the most translucent case) and \mbox{$x$($^{12}$CO)\,=\,(0.5-1.5)$\cdot$10$^{-4}$} otherwise.

In order to determine $M_{\rm g}$, we use the observed angular sizes of each globule. 
Table~\ref{table:LTE-globules} summarizes the radius ($R_{\rm g}$) of each globule, and the range of gas densities ($n_{\rm H}$\,$\simeq$\,$N_{\rm H}$\,/\,2$R_{\rm g}$) and molecular core depths (defined as $A_{\rm V}$\,=\,3.5$\cdot$10$^{-22}$\,$N$($^{12}$CO)/$x$($^{12}$CO), see Sect.~\ref{sec:PDR-mods}) derived in the single-slab approximation.

In this approach, the most critical error parameter in the calculated values is the adopted range of CO abundances. The adopted $x$($^{12}$CO) range (within a factor of three) roughly agrees with specific PDR models adapted to the UV illuminating conditions in the shell  (see Sect.~\ref{sec:PDR-mods}). The resulting range of calculated  $A_V$ values in each globule brackets the extinction values we obtain from the more detailed PDR models as well. They are also consistent with the nondetection of C$^{18}$O emission, which approximately implies $A_V$\,$<$\,3\,mag for all globules (except for the globules \#5 and \#10).

Taking into account the achieved rms sensitivity of our maps, we also computed  the minimum beam-averaged $^{12}$CO column density, 
\mbox{$N_{\rm min}$($^{12}$CO)},  that we could have detected. We adopt  \mbox{$W_{3\sigma}=3\sigma \sqrt{2\,\delta {\rm v}\Delta {\rm v}}$}                
with \mbox{$\sigma$=0.25\,K} (the rms noise level of our native map), 
$\delta$v=0.25\,km\,s$^{-1}$ (the velocity channel resolution), 
$\Delta$v=2\,km\,s$^{-1}$ (the expected line-width). 
This leads to $W_{3\sigma}$=0.75\,K\,km\,s$^{-1}$. 
For $T_{\rm ex}$=5-20\,K, our map is sensitive to a beam-averaged \mbox{$N_{\rm min}$($^{12}$CO)\,$>$\,(5-15)$\cdot$10$^{14}$\,cm$^{-2}$} (3$\sigma$). Assuming a typical shell thickness of 1.8\,mag of visual 
extinction in the Veil \citep{Odell01}, this limit is equivalent to
detecting  $x$($^{12}$CO) abundances above 
(1-3)$\cdot10^{-7}$ in the shell. This threshold seems high, but we highlight the strong UV illumination conditions and low extinction depth  of this foreground component (i.e., low molecular column densities). 
On the other hand, the $N$($^{12}$CO) column we  detect toward the  \mbox{negative-v$_{\rm LSR}$} globules is at least 25 times higher than 
$N_{\rm min}$($^{12}$CO). This implies that widespread and abundant
CO is largely absent from the shell.

\begin{table*}[t]
\caption{Velocity dispersions, pressures, and relevant masses of each globule molecular core (from $^{12}$CO and $^{13}$CO) and their envelopes (from C$^+$). \label{table:globule-masses}} 
\centering \small 
\begin{tabular}{rccccccccc@{\vrule height 8pt depth 5pt width 0pt}}
\hline\hline
       &  $\sigma_{\rm nth}$(C$^+$) & $P_{\rm nth}/k$\,(C$^+$)$^a$ & $P_{\rm th}/k$\,(C$^+$)$^a$ & $\sigma_{\rm nth,\,g}$$^b$ & $P_{\rm nth,\,g}/k$$^b$ & $P_{\rm th,\,g}/k$\,($T_{\rm ex}$)$^c$ & $M_{\rm g}$  & $m_{\rm BE}$          & $M_{\rm J}$\\
 \small{globule} &    (km\,s$^{-1}$)$^a$      & (10$^6$\,cm$^{-3}$\,K)   & (10$^6$\,cm$^{-3}$\,K) &  (km\,s$^{-1}$)  & (10$^6$\,cm$^{-3}$\,K) &  (10$^6$\,cm$^{-3}$\,K)    & ($M_{\odot}$)    & ($M_{\odot}$)$^{c,d}$ & ($M_{\odot}$)$^{e}$\\\hline
  \#1  & 2.3(0.3) & 42.2$\pm$24.5 & 4.8$\pm$2.8  & 0.5(0.2) & 3.8$\pm$2.2 & 0.4$\pm$0.2  & 0.08$\pm$0.05 &  0.03$\pm$0.01 &  3.0(2.4) \\  
  \#2  & 2.8(0.2) & 33.0$\pm$19.8 & 2.5$\pm$1.5  & 0.5(0.1) & 1.8$\pm$1.1 & 0.3$\pm$0.2  & 0.3$\pm$0.2  &  0.06$\pm$0.03 &  5.1(3.4)\\ 
  \#3  & 1.0(0.1) & 5.3$\pm$2.8   & 3.0$\pm$1.6  & 0.5(0.1) & 1.8$\pm$1.0 & 0.5$\pm$0.3  & 0.2$\pm$0.1  &  0.2$\pm$0.1 &  3.6(0.6)\\
  \#4  & 1.0(0.1) & 3.5$\pm$1.8   & 2.1$\pm$1.1  & 0.5(0.1) & 1.5$\pm$0.8 & 0.5$\pm$0.3  & 0.4$\pm$0.2  &  0.6$\pm$0.3 &  6.3(1.0)\\
  \#5  & 2.5(0.1) & 117.4$\pm$65.6 & 10.9$\pm$6.1 & 0.9(0.1) & 22.1$\pm$12.4 & 5.4$\pm$3.5  & 1.5$\pm$0.8  &  0.7$\pm$0.4 & 15.5(1.5)\\
  \#6  & 1.1(0.1) & 9.4$\pm$5.4   & 4.3$\pm$2.5  & 0.5(0.2) & 3.3$\pm$1.9 & 0.4$\pm$0.3  & 0.05$\pm$0.03 &  0.06$\pm$0.03 &  2.4(1.2)\\
  \#7  & 1.5(0.1) & 15.6$\pm$8.7  & 4.3$\pm$2.4  & 0.4(0.1) & 2.4$\pm$1.4 & 2.1$\pm$1.4  & 0.3$\pm$0.2  &  1.8$\pm$0.9 &  3.7(2.0)\\
  \#8  & 1.2(0.2) & 2.2$\pm$1.2   & 0.9$\pm$0.5  & 0.4(0.1) & 0.4$\pm$0.2 & 0.3$\pm$0.2  & 1.2$\pm$0.7  &  2.5$\pm$1.2 & 8.4(5.1)\\
  \#9  & 1.0(0.1) & 1.6$\pm$0.9   & 0.9$\pm$0.5  & 0.6(0.1) & 0.8$\pm$0.4 & 0.3$\pm$0.2  & 0.11$\pm$0.06 &  2.4$\pm$1.2 &  7.0(1.0)\\
 \#10  & 2.0(0.1) & 33.4$\pm$19.2 & 4.7$\pm$2.7  & 0.5(0.1) & 3.7$\pm$2.1 & 3.0$\pm$2.0  & 4.7$\pm$2.7  &  2.1$\pm$1.1 & 12.3(1.9)\\
  \hline
\tiny{Median} & 1.3        & 12.5          & 3.6          &  0.5               &  2.1  & 0.4  & 0.3   & 0.6    & 5.6\\
  \hline

\end{tabular}
\tablefoot{Based on the range of globule densities estimated in Table~\ref{table:LTE-globules}. Values in parenthesis are 1$\sigma$ errors. 
$^a$Assuming $T_{\rm k}$\,=\,100\,K. $^b$From \mbox{$^{13}$CO (2-1)} line-widths. $^c$Assuming $T_{\rm k}$\,=\,$T_{\rm ex}$($^{12}$CO).
$^d$\mbox{Bonnor-Ebert} mass for an external pressure given by \mbox{$P_{\rm ext}$\,=\,$P_{\rm nth}$(C$^+$)\,+\,$P_{\rm th}$(C$^+$)}.
$^e$Jeans mass for $R_{\rm g}$ and $\sigma_{\rm nth+th,\,g}$ determined from $^{13}$CO~(2-1).}
\end{table*}

\subsection{Globule velocity dispersions, pressures, and masses}\label{sec:masses}

Table~\ref{table:globule-masses} summarizes the properties of the interior of each globule (as traced by $^{13}$CO) and of their envelope and/or surroundings (as traced by C$^+$). This table displays the nonthermal (turbulent) and thermal gas pressures, the estimated
\mbox{Bonnor-Ebert} mass for a pressure-confined isothermal sphere 
\citep[$m_{\rm BE}$;][]{Ebert55,Bonnor56}, and the Jeans mass 
\citep[$M_{\rm J}$; e.g.,][]{Larson78}. To derive these masses, we assume
$T_{\rm k}$=$T_{\rm ex}$($^{12}$CO\,2-1) and use: 
\begin{equation}  
m_{\rm BE}=1.15\, \left(\frac{c_s}{0.2\,{\rm km\,s^{-1}}}\right)^4 
  \left( \frac{P_{\rm ext}} {10^5 {\rm cm^{-3}\,K}}    \right)^{-0.5},
\end{equation} 
from \cite{Lada08},  where $c_s$  is the temperature-dependent speed of sound inside the  globule. For the Jeans mass we use:
\begin{equation} 
M_{\rm J}= \frac{5\,R_{\rm g}\, \sigma^2} {2G}, 
\end{equation} 
from \cite{Kirk17}, where $\sigma$ includes the dominant \mbox{nonthermal} support inside each globule (from $^{13}$CO line-widths). The resulting median values of the sample  are $M_{\rm g}$\,=\,0.3\,$M_{\odot}$, $m_{\rm BE}$\,=\,0.6\,$M_{\odot}$, and 
$M_{\rm J}$\,$=$\,5.6\,$M_{\odot}$ (see Table~\ref{table:globule-masses} individually).

\subsection{Stellar FUV photon flux ($G_0$) toward the shell}\label{sec-G0}  

\begin{figure}[t]
\centering   
\includegraphics[scale=0.165, angle=0]{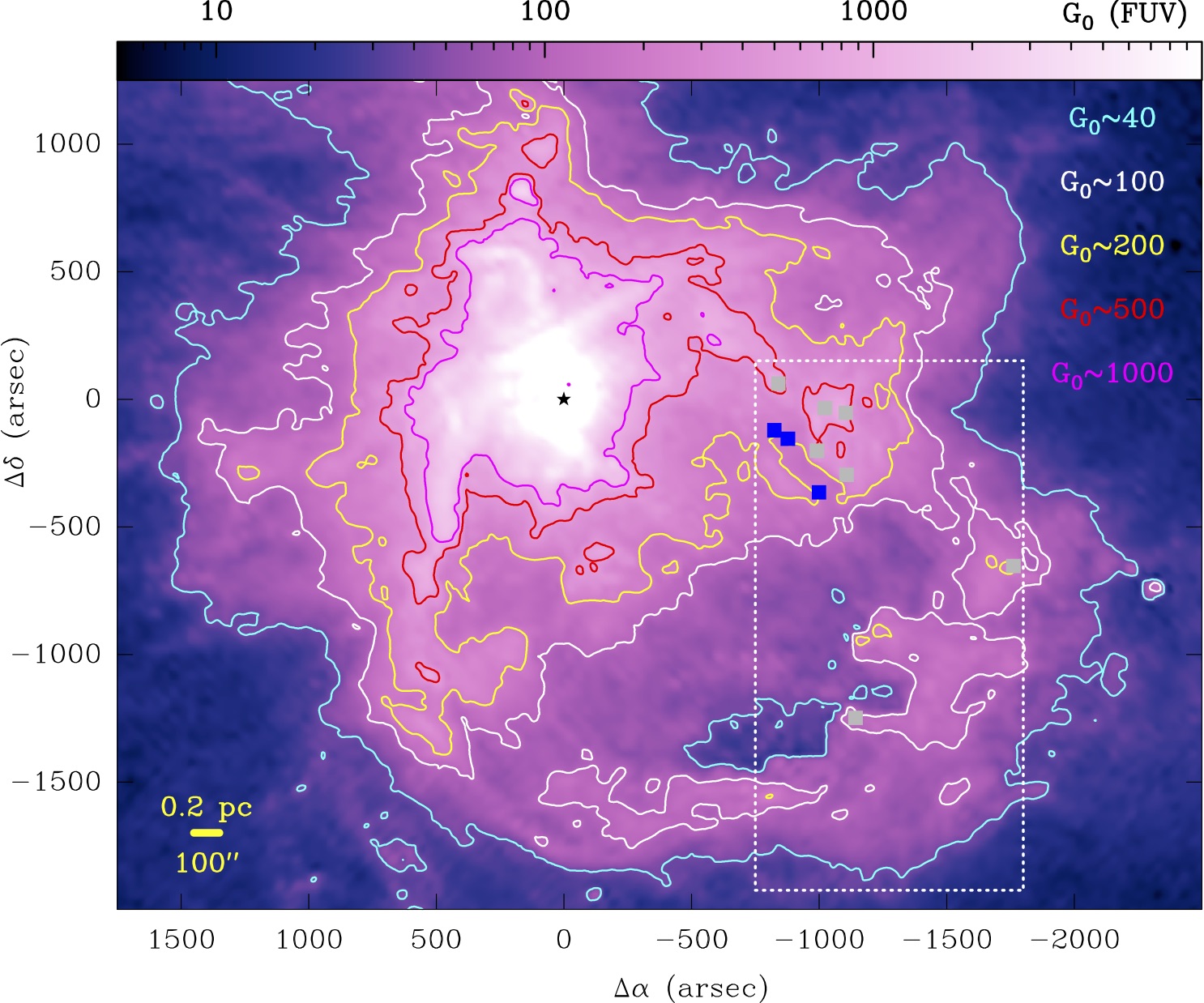}
\caption{One-square-degree image of the flux of nonionizing FUV photons ($G_0$  in units of the Habing field) along lines of sight toward the Veil bubble. The black star at $(0'',0'')$ corresponds to the position of the illuminating star 
\mbox{$\theta^1$ Ori C}. Here, we focus on the  area enclosed by the dotted box. The blue and gray squares mark the position of the detected CO globules (blue squares for the \mbox{negative-v$_{\rm LSR}$} globules).}\label{fig:pacs70}
\end{figure}

In \cite{Goico15,Goico19} we estimated $G_0$ in \mbox{OMC-1} from the integrated far-IR (FIR) dust thermal emission observed
by Herschel.  When dust grains absorb FUV photons, they are heated up, and re-radiate at FIR wavelengths. For a face-on PDR:   
\begin{equation}
G_0 \simeq \frac{1}{2}  \frac{I_{\rm FIR} (\rm erg\,s^{-1}\,cm^{-2}\,sr^{-1})}{1.3\cdot 10^{-4}},
\end{equation}
from \cite{Hollenbach97}, where $G_0$ is the FUV radiation field in Habing units 
\citep[1.6$\cdot$10$^{-3}$\,erg\,s$^{-1}$\,cm$^{-2}$;][]{Habing68}, and 
$T_{\rm d,\,PDR}\simeq 12.2\,G_{0}^{0.2}$ is a characteristic dust temperature
in the PDR \citep{Hollenbach91}. In practice, the longer wavelength submillimeter (submm) dust emission toward the lines-of-sight of large column density may not only be produced by \mbox{FUV-heated} grains but may have a contribution from colder dust in the background molecular cloud. The emission from FUV-irradiated warm dust is more easily detected at shorter FIR wavelengths. Indeed, the shell morphology in the PAH\,8\,$\upmu$m  and PACS\,70\,$\upmu$m emission  is very similar and nicely delineates the expanding shell.  However, the bubble morphology is less apparent at longer submm wavelengths. Hence, to create an approximate map of $G_0$ along the line of sight toward the shell, shown in Fig.~\ref{fig:pacs70}, we used:
\begin{equation}
 {\rm log}_{10}\,G_0 = (0.975\pm0.02)\,{\rm log}_{10}\,I_{70} - (0.668\pm0.007),
\end{equation} 
  where $I_{70}$ is the 70\,$\upmu$m dust surface brightness in MJy\,sr$^{-1}$. We obtained this scaling after determining $G_0$ from SED fits toward the irradiated  surface of \mbox{OMC-1} \citep[][]{Goico15}. Because photometric observations detect the dust continuum emission projected in the plane of the sky, it is not easy to resolve the dust temperature (or $G_0$) gradient along each line of sight. Hence, the $G_0$ contours shown in Fig.~\ref{fig:pacs70} should be understood as the maximum FUV flux  that can impinge a globule located in a given position sightline. The minimum value $G_0$\,$\simeq$\,40 (or $T_{\rm d,\,PDR}$\,$\simeq$\,25\,K) is representative of the
most distant shell edges far from the Trapezium. This $G_0$ value likely   represents
the local FUV flux around the  \mbox{negative-v$_{\rm LSR}$} globules.

\begin{table}[t]
\caption{Parameters used in the PDR models of the \mbox{negative-v$_{\rm LSR}$} 
globules  and comparison with observed intensities.\label{table:PDR-mods}} 
\centering
\begin{tabular}{ccc@{\vrule height 8pt depth 5pt width 0pt}}
\hline\hline
Model parameter                                 &     Value                                    &     Note        \\ 
\hline
Local $G_0$                                                     &     40 Habing                               &    Best model   \\
Total depth $A_{\rm V}$                         &     1.8 mag                                      &                 \\
Gas density $n_{\rm H}$                         &  2$\cdot$10$^4$\,cm$^{-3}$                  &                 \\
\hline
Cosmic Ray $\zeta_{\rm CR}$                     & 10$^{-16}$\,H$_2$\,s$^{-1}$                       &   $a$           \\
$R_{\rm V}$\,=\,$A_{\rm V}$/$E_{\rm B-V}$       &          5.5                                 &  Orion$^b$      \\
$N_{\rm H}$/$A_{\rm V}$                         & 2.86$\cdot$10$^{21}$\,cm$^{-2}$\,mag$^{-1}$ &  Orion$^b$      \\
$M_{\rm gas}/M_{\rm dust}$                      & 100                                         &  Local ISM      \\
Abundance O\,/\,H                               & 3.2$\cdot$10$^{-4}$                         &                  \\
Abundance $^{12}$C\,/\,H                                                & 1.5$\cdot$10$^{-4}$                                   &  Orion$^c$      \\
Abundance N\,/\,H                                                               & 7.5$\cdot$10$^{-5}$                                   &                                   \\
Abundance S\,/\,H                                                           & 1.5$\cdot$10$^{-5}$                                      &                                   \\
$^{12}$C\,/\,$^{13}$C                                       &    67                                       &  Orion$^d$      \\
\hline
\hline
Species                                         &  Predicted $N$(cm$^{-2}$)                   &                  \\ 
\hline
C$^+$                                           &   3.5$\cdot$10$^{17}$                       & Best model \\
$^{12}$CO                                                                               &   5.2$\cdot$10$^{15}$                                 & Best model \\
$^{13}$CO                                       &   1.8$\cdot$10$^{14}$                       & Best model \\
\hline
                                                                                                &          Predicted $W$                          & Observed \\
Line                                            &   (K\,km\,s$^{-1}$)                         & (K\,km\,s$^{-1}$)\\
\hline   
[\CII]\,158$\upmu$m                             &       32.3                                  & 29.3$^\dagger$$\pm$6.0$^\ddagger$\\
$^{12}$CO (2-1)                                                                 &       10.8                                                                & 9.5$^\dagger$$\pm$4.3$^\ddagger$\\
$^{13}$CO (2-1)                                                                 &        0.4                                                            & 0.7$^\dagger$$\pm$0.3$^\ddagger$\\
\hline                                    
\end{tabular}
\tablefoot{$^a$From \cite{Indriolo15}. $^b$From \citet{Lee68} and \citet{Cardelli89}. $^c$From \cite{Sofia04}. $^d$From \cite{Langer84}.
$^\dagger$Mean line intensities toward globules \#1, \#2, and \#3. $^\ddagger$Standard deviation.}
\end{table}

\begin{figure*}[h]
\centering   
\includegraphics[scale=0.415, angle=0]{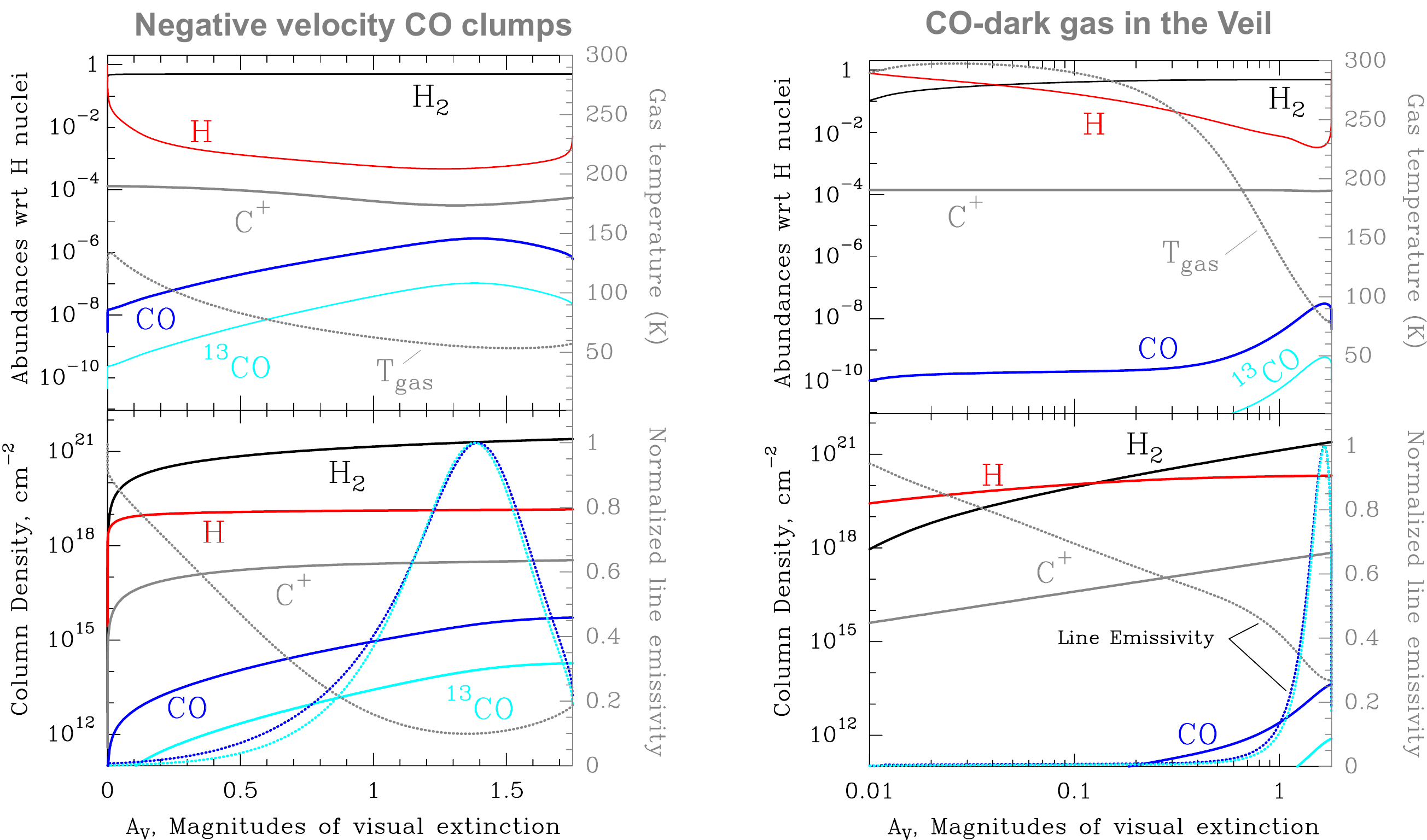}
\caption{Profiles of predicted abundance, gas temperature, column density, and line emissivity (dotted curve) as a function of globule depth.  \mbox{\textit{Left panel}}: Model that reproduces the observed [\CII]\,158\,$\upmu$m, \mbox{$^{12}$CO} and \mbox{$^{13}$CO (2-1)} line intensities toward the \mbox{negative-v$_{\rm LSR}$} CO globules (\#1, \#2, and \#3). The FUV field is \mbox{$G_0$\,=\,40} and the gas density is 
\mbox{$n_{\rm H}$\,=\,2$\cdot$10$^4$\,cm$^{-3}$}. \textit{Right panel:} Model appropriate for a representative position of the shell where CO is not detected, with \mbox{$n_{\rm H}$\,=\,3$\cdot$10$^3$\,cm$^{-3}$}, 
\mbox{$G_0$\,=\,200}, and \mbox{$A_{\rm V}$\,=\,1.8}
\citep[from][and references therein]{Abel19}. 
}\label{fig:pdr_mods}
\end{figure*}

\subsection{Depth-dependent globule photochemical models}\label{sec:PDR-mods}

In this section we go beyond the single-slab analysis and model
 the possible abundance and temperature gradients across a \mbox{FUV-irradiated} 
 globule. We use the Meudon PDR code \mbox{\citep{LePetit06}} to model the penetration of FUV radiation \citep{Goico07}, thermal balance \citep{Bron14}, steady-state gas chemistry, and  \mbox{NLTE [\CII]\,158\,$\upmu$m}, $^{12}$CO, and $^{13}$CO excitation 
 and radiative transfer \citep{Gonzalez08}. 
 We adopted an extinction-to-color-index ratio, 
\mbox{$R_{\rm V}=A_{\rm V}/E_{\rm B-V}$}, of 5.5 \citep[consistent with the flatter extinction curve observed in the material toward the Trapezium stars in Orion,][]{Lee68,Cardelli89}, and a \mbox{$A_{\rm V}$/$N_{\rm H}$} ratio of 3.5$\cdot$10$^{-22}$\,mag\,cm$^2$ appropriate for the material in Orion's Veil \citep[e.g.,][]{Abel16,Abel19}. The adopted model parameters and elemental abundances are tabulated in
Table~\ref{table:PDR-mods}.

 The chemical network includes specific 
$^{13}$C isotopic fractionation reactions \citep{Langer84}  meaning that the $^{13}$CO predictions are accurate. In FUV-irradiated gas reaction 
\mbox{$^{13}$C$^+$ + CO $\rightleftarrows$ C$^+$ + $^{13}$CO + $\Delta E$}~(1)  
\citep[where $\Delta E$=35\,K is the zero-point energy difference between the species on either side of the reaction,][]{Watson76}  transfers $^{13}$C$^+$ ions and makes $^{13}$CO more abundant than the $^{12}$CO/$R_{\rm 12C13}$  ratio if $T_{\rm k}$ reaches $\sim$50\,K and below   
\citep[e.g.,][]{Rollig13}. 

In our models, one side of the globule is illuminated by 
$G_0$\,$\geq$\,40. 
The other side of the globule is illuminated by $G_0$\,=\,1.7. 
Starting around
the extinction depths and column densities estimated from the single-slab analysis, we varied the local flux of FUV photons impinging the globule, the gas density $n_{\rm H}$ (first assumed to be constant), and $A_{\rm V}$ (the depth into the CO globule in magnitudes of visual extinction), and tried to fit the observed [\CII]\,158\,$\upmu$m, \mbox{$^{12}$CO (2-1)}, and \mbox{$^{13}$CO (2-1)} line intensities. Best model parameters and line emission predictions for the \mbox{negative-v$_{\rm LSR}$} globules are shown in Table~\ref{table:PDR-mods}.

The best-fit parameters to the observed line emission from the \mbox{negative-v$_{\rm LSR}$} globules (\#1, \#2, and \#3) are: $G_0$\,$\approx$\,40, \mbox{$n_{\rm H}$\,$\simeq$\,2$\cdot$10$^4$\,cm$^{-3}$}, and $A_{\rm V}$\,$\simeq$\,1.8\,mag, giving a spatial size of  \mbox{2.5$\cdot$10$^{17}$\,cm\,=\,\,8,300\,AU\,$\simeq$\,2\,$R_{\rm g}$}. This length scale approximately agrees with the observed radii of these globules. 
We  note that an isobaric model (constant thermal pressure and varying density)
with \mbox{$P_{\rm th}/k$\,$\simeq$10$^6$ cm$^{-3}$\,K} and  $G_0$\,$\simeq$\,50 predicts roughly the same  intensities as the above constant density model. 
  This thermal pressure is a good compromise between the values of $P_{\rm th}/k$\,(C$^+$) and $P_{\rm th, g}/k$ (the internal globule pressure) inferred in the single-slab approximation (Table~\ref{table:globule-masses}).
 We note that $G_0$ in these models is the \textit{local} FUV flux around the globule needed to reproduce the observed line intensities, whereas the  $G_0$ map in Fig.~\ref{fig:pacs70} shows the integrated FUV flux along the line of sight.  
 The lower local $G_0$ values are consistent with the fact that these globules are embedded in neutral gas  and dust that attenuates the FUV photon flux irradiating  the shell.

     The abundance, gas temperature, column density, and line emissivity profiles predicted by the  constant density model are shown in Fig.~\ref{fig:pdr_mods} (left). 
\mbox{Figure~\ref{fig:pdr_mod_inten}} specifically compares the predicted  and observed line intensities for the \mbox{negative-v$_{\rm LSR}$} globules individually.
    These, and also the small globule \#6 (which is close in LSR velocity and shows similar emission properties), can be fitted within a factor three of the estimated local $G_0$ value (40), and within a 
    \mbox{factor three} of the estimated gas density ($n_{\rm H}$\,=\,2$\cdot$10$^4$\,cm$^{-3}$). 
This density is up to \mbox{$\sim$10-40} times higher than the density in the most common portions of the expanding shell, which do not show CO emission \citep[see][for their component III(B) of the Veil associated with the neutral shell]{Abel19}.

   As observed, our PDR models of the  \mbox{negative-v$_{\rm LSR}$} globules predict that the C$^+$/CO abundance ratio throughout the globule is  above one, with a maximum abundance $x$(CO) of several 10$^{-6}$. Also, as observed,  the model predicts an enhancement of the $^{13}$CO column density produced by isotopic fractionation. Indeed, close to the CO abundance peak (at about $A_{\rm V}$\,=\,1.5\,mag  for this combination of $G_0$ and $n_{\rm H}$ values) the gas temperature drops to $\sim$50\,K and reaction (1) becomes the dominant formation route for $^{13}$CO. This chemical effect favors the overproduction of $^{13}$CO and explains the detected \mbox{$^{13}$CO (2-1) emission from the globules} but no [$^{13}$\CII] emission (\mbox{S. Kabanovic}, Priv. Comm.). 

   A few of the brighter positive-v$_{\rm LSR}$ globules  show much brighter 8\,$\upmu$m, [\CII]\,158\,$\upmu$m, and \mbox{$^{12}$CO~(2-1)} emission levels \mbox{(Table~\ref{table:line-fits})}, up to $\sim$100\,K\,km\,s$^{-1}$  (see globules \#5 and \#10 and their  8\,$\upmu$m and [\CII]\,158\,$\upmu$m bright rims). These are compatible with PDR models of higher FUV irradiation doses ($G_0$\,$\gtrsim$\,500, i.e., 
globules that are more exposed to the unattenuated radiation field from the Trapezium), higher gas densities ($n_{\rm H}$\,$\gtrsim$\,4$\cdot$10$^4$\,cm$^{-3}$), and higher extinction depths ($A_{\rm V}$\,$\gtrsim$\,5\,mag). These higher $n_{\rm H}$ and $A_{\rm V}$ values are consistent with our detection of \mbox{C$^{18}$O~(2-1)} line emission only toward globules \#5 and \#10. 

   Finally, the right panels in Fig.~\ref{fig:pdr_mods}  are PDR models for  a representative position in the Veil shell where CO is not detected. We chose $G_0$\,=\,200, $n_{\rm H}$\,=\,3$\cdot$10$^3$\,cm$^{-3}$ and $A_{\rm V}$\,=\,1.8\,mag 
    \citep[e.g.,][]{Abel16,Abel19}. The predicted C$^+$ column density (\mbox{$N$(C$^+$)\,$\simeq$\,10$^{18}$\,cm$^{-2}$}) and [\CII]\,158\,$\upmu$m line intensity ($\sim$50\,K\,km\,s$^{-1}$) are in line with the values typically observed across the shell
\citep{Pabst19,Pabst20}. However, the predicted CO column density is very low,  
$N$(CO)\,$\simeq$\,4$\cdot$10$^{13}$\,cm$^{-2}$ (most carbon is in the form of C$^+$),  and the expected \mbox{$^{12}$CO (2-1)} line intensity ($\sim$0.08\,K\,km\,s$^{-1}$) is too faint to be detected. We conclude that owing to the  strong stellar FUV irradiation conditions, most of the translucent gas  in the shell  will typically have exceedingly low CO column densities.

\section{Discussion}

\subsection{Fraction of the shell mass traced by CO globules}

\begin{figure*}[]
\centering   
\includegraphics[scale=0.21, angle=0]{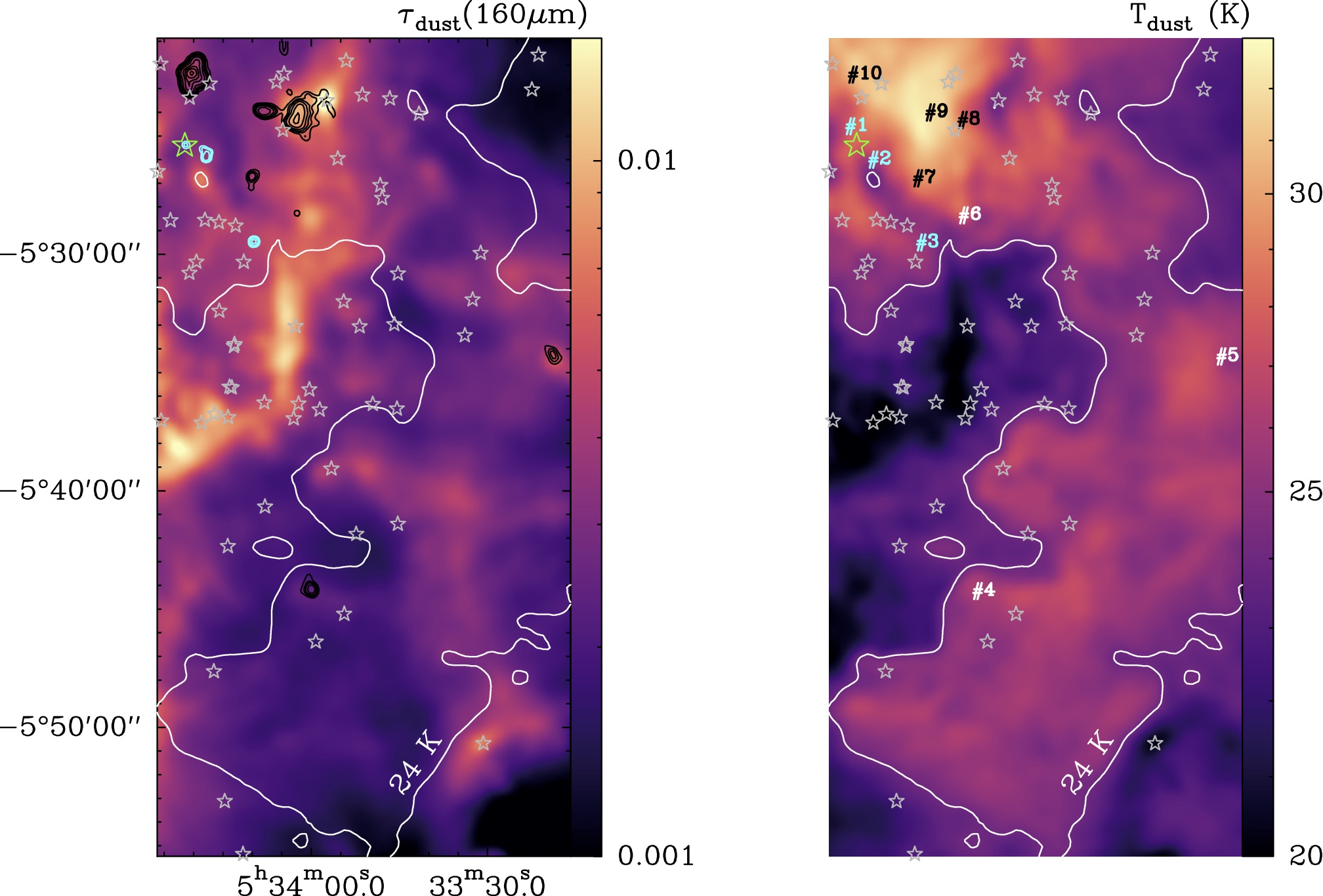}
\caption{Dust opacity at 
160\,$\upmu$m (left panel) and the dust temperature (right) computed by fitting  modified blackbodies to the continuum emission measured by Herschel \citep{Pabst19}. The white contour at  $T_{\rm d}$=24\,K roughly encloses the expanding shell as observed in 
[\CII]\,158\,$\upmu$m and PAH emission. The lower-$T_{\rm d}$ regions outside this area mostly arise from the background molecular cloud and not from the shell. The left panel explicitly shows the \mbox{$^{12}$CO (2-1)} emission from the globules in cyan (black) contours for the negative (positive) LSR velocities. The right panel shows their labels. In both maps, the gray stars show the position (in the plane of the sky) of YSOs previously detected from infrared observations \citep{Megeath12,Megeath16}. Only globule \#1 matches the position of a YSO (object \#1728) shown with a green star.}\label{fig:dust}
\end{figure*}

In \cite{Pabst19,Pabst20}, we estimated  the total mass of the expanding half-shell 1,500\,-\,2,600\,$M_{\odot}$ (from the FIR dust opacity and from the 
[\CII]\,158\,$\upmu$m emission itself). The region studied in the current work is about one-third of the shell and we determine that the shell mass in this area of interest is  \mbox{$\sim$400\,-\,700\,$M_{\odot}$}. This is the mass enclosed  inside the $T_{\rm d}$\,$>$\,24\,K contour shown in Fig.~\ref{fig:dust} (right panel). Here, $T_{\rm d}$ is an effective temperature obtained from a modified black body fit to the 70, 100, 160, 250, 350, and 500\,$\upmu$m photometric emission measured by Herschel \citep{Andre10}.  The morphology of the area with $T_{\rm d}$\,$>$\,24\,K resembles that revealed by the 
[\CII]\,158\,$\upmu$m, PACS\,70, and PAH\,8\,$\upmu$m images that delineates the expanding shell. All  CO globules lie inside the 
$T_{\rm d}$\,$>$\,24\,K area. We compute the mass of the CO globules by considering their observed sizes (as seen in $^{12}$CO) and their column densities determined from the single-slab analysis. We derive a total mass that ranges from 4\, to 14\,$M_{\odot}$ adding their  masses.
Therefore,  they only account for $<3\,\%$ of the shell mass.

\subsection{Magnetic support and external pressure confinement}

The CO globules  have the following median properties:
radius $R_{\rm g}$\,$\simeq$\,7,100\,AU,
gas density \mbox{$n_{\rm H}$\,=\,3.6$\cdot$10$^4$\,cm$^{-3}$}, and
mass \mbox{$M_{\rm g}$\,=\,0.3\,$M_{\odot}$}.  The observed $^{13}$CO line profiles 
(Fig.~\ref{fig:spectra} and Table~\ref{table:line-fits}) imply supersonic  nonthermal motions inside them, with a median turbulent velocity dispersion of  $\sigma_{\rm turb}$($^{13}$CO)\,=\,0.5\,km\,s$^{-1}$. Turbulent pressure dominates thermal pressure in the globules (see Table~\ref{table:globule-masses} for  each globule individually). The [\CII]\,158\,$\upmu$m line profiles are systematically broader, indicating
more extended gas flows, as well as higher turbulent velocity dispersions in their envelopes  and in the interglobule medium, with a median of 
$\sigma_{\rm turb}$(C$^+$)=1.3\,km\,s$^{-1}$. This value implies high turbulent pressures of \mbox{$P_{\rm turb}$(C$^+$)\,=\,$\rho$\,$\sigma_{\rm turb}^{2}$} (with $\rho$\,=\,$\mu\,m_{\rm H}\,n_{\rm H}$) around the CO globules. 
Their external pressure \mbox{(thermal plus turbulent)} is 
$P_{\rm ext}$/$k$\,$\gtrsim$10$^7$\,cm$^{-3}$\,K (median),  higher than their internal pressures ($P_{\rm g}$/$k$ of a few 10$^6$\,cm$^{-3}$\,K).

Velocity-resolved HI  Zeeman observations toward thin, low-density ($n_{\rm H}$\,$\gtrsim$\,10$^3$\,cm$^{-3}$)   portions of the Veil have suggested typical (line-of-sight) magnetic field strengths of \mbox{$B_{\rm los}$\,=\,$50$-$75$\,$\upmu$G},
reaching  \mbox{$B_{\rm los}$\,=\,100\,$\upmu$G} or more
toward several positions \mbox{\citep{Troland16}}.
This latter value implies magnetic pressures of \mbox{$P_{\rm B}$/$k$\,=\,$B_{\rm tot}^{2}/8\pi k$\,$\geq$\,8$\cdot$10$^6$\,K\,cm$^{-3}$} 
 if   \mbox{$B_{\rm los}^{2}$\,=\,$B_{\rm tot}^{2}/3$}  \mbox{\citep{Crutcher99,Abel16}}.
 Hence, the Veil seems to be a very magnetized medium, with the energy in $B_{\rm tot}$  being similar to or more than that in gas motions or gravity.
In general,  $B_{\rm tot}$ increases with density at ionized and molecular gas boundaries
\citep[\mbox{$B_{\rm tot}\propto n_{\rm H}^{0.5-1}$}, e.g.,][]{Planck16}. Hence, we can expect stronger fields threading the denser  CO globules. As an example, the derived intensity of the (plane-of-the-sky) magnetic field 
in the Orion Bar PDR is $\sim$300\,$\upmu$G \citep[derived from the polarized FIR dust emission;][]{Chuss19}. In the swept-up material that confines the bubble, compression associated with the expansion of the shell itself will further increase the magnetic field as, given the high degree of ionization in PDR gas, \mbox{$n$($e^-$)\,$\approx$\,$n$(C$^+$)}, the field lines will be \mbox{``frozen-in''}. 
Overall, we conclude that plausible magnetic fields 
($B_{\rm los}$\,$\approx$100-200\,$\upmu$G)  would be sufficient to magnetically support
 the CO globules. We note that  simulations do predict that FUV-irradiated cometary globules can be magnetically supported \citep[e.g.,][]{Lefloch94,Henney09,Mackey11}.

The expanding shell is certainly  massive, \mbox{1,500-2,600\,$M_{\odot}$} \citep{Pabst20}, comparable to the current mass of the molecular core behind 
\citep[\mbox{OMC-1}; e.g.,][]{Genzel89,Bally87}. The shell material has been swept up by the disruptive effects of  \mbox{$\theta^1$ Ori C}  winds and UV radiation.
However, in the surveyed area, the mass of CO globules only represents $<$\,3\% of the total shell mass. Assuming $G_0$\,$>$\,40, the timescale for photodissociation of CO in the shell is considerably shorter than its expansion timescale  \citep[\mbox{$\tau_{\rm exp}$\,$\simeq$\,200,000\,yr};][]{Pabst19}. 
This is consistent with the lack of widespread and extended CO emission in the Veil and leaves the origin and fate of the detected molecular globules as an open question. We discuss this issue, at least qualitatively, in the following section.

\vspace{-0.2cm}
\subsection{Origin and evolution of the CO globules in the Veil}

Globules may form in-situ by hydrodynamic instabilities \citep[see e.g.,][]{Schneps80,Sharp84,Murray93,Nakamura06}
     at the interface between the (shell's) neutral gas that rests on a light and rarefied (about 1\,cm$^{-3}$) hot plasma \citep{Gudel08}.  This interface is \mbox{Rayleigh-Taylor} unstable \citep[][]{Spitzer54} and could form ``trunks'' or ``fingers'' of size $R$ on timescales of \mbox{$\tau_{\rm RT}$\,$\approx$\,$R$/$c_s$\,$\simeq$\,10$^5$\,($R$\,/\,0.1\,pc)\, yr}
 (where $c_s$ is the speed of sound in the shell: 1\,km\,s$^{-1}$ at 100~K). The observed sizes of the CO globules, about 10\% of the total shell thickness, indeed imply that they could form during the expansion of the shell into an environment of lower density \citep[\mbox{$\tau_{\rm RT}$\,$<$\,$\tau_{\rm exp}$}, see e.g.,][]{Schneps80}. \mbox{Kelvin-Helmholtz} instabilities could also develop at the side of the fingers over similar timescales \citep{Sharp84,Murray93,Nakamura06,Berne12}. Indeed, \mbox{wave-like} emission structures (typical of these instabilities) with spatial wavelengths ranging from 0.1 to 0.01\,pc have been observed at several \mbox{ionized--molecular} gas interfaces in OMC-1. In particular, \citet{Berne10} and \citet{Berne12} analyzed the \mbox{``KH ripples''},  with \#7 being the head of this structure  (see Fig.~\ref{fig:globule-images} for a detailed view). Much higher
resolution ALMA images of the Orion Bar PDR 
suggest smaller scale density undulations separated by $\sim$0.01\,pc at the  
\mbox{FUV-irradiated} boundary between
\mbox{OMC-1} and the \mbox{Huygens} \HII~region \citep{Goico16}.

Alternatively, the CO globules could have been pre-existing molecular structures \citep[e.g., like in][]{Reipurth83}, which are denser than the shell, and are perhaps generated by the turbulent velocity field in \mbox{the OMC} \citep[e.g.,][]{Hartman07,Hacar17}. These globules are compressed and swept along by a passing shock accompanying the expanding shell \citep[e.g.,][]{PikelNer74}.  
 If the initial gas density contrast between globules and shell has a value of about 10, the shock wave penetrating the CO globules will be slower by a factor of approximately three. This gives a shock velocity of v$_{\rm shock}$\,$=$\,4\,km\,s$^{-1}$, fast enough to traverse the globule in  \mbox{$\tau_{\rm shock}$\,$\approx$\,$R_{\rm g}$\,/\,v$_{\rm shock}$\,$\simeq$\,10,000\,yr} and trigger compression. Gas cooling through CO lines will decrease the post-shock gas temperature and further enhance gas compression and density. 
This could result in free-fall times, $t_{ff}$\,$\approx$\,$(G\,\rho)^{-1/2}$, comparable to the shell expansion timescale ($\tau_{\rm exp}$)  if globules are compressed to densities \mbox{$>$2$\cdot$10$^4$\,cm$^{-3}$} (comparable with our estimated values). However,
given the physical conditions in the Veil shell: enhanced turbulence, FUV-heating,
and strong magnetic field, it may be unlikely that gravity can beat the pressure support of the globule (see further discussion in the following section).

The detected CO globules are surrounded by larger structures
of  predominantly neutral gas and dust, emitting thermal continuum at 70\,$\upmu$m, PAHs,  and [\CII]\,158\,$\upmu$m (that matches the velocity of the CO emission; see Fig.~\ref{fig:globule-images-Cplus}). This material likely shields the  globules from  extreme UV (EUV; $E$\,$>$13.6\,eV) ionizing radiation and reduces the flux of dissociating FUV photons reaching their interiors. 
This scenario is consistent with the observed low CO excitation temperatures ($\sim$10\,K) and moderate \textit{local}  $G_0$ values around the \mbox{negative-v$_{\rm LSR}$} globules (and \#6). It is also consistent with their apparent spherical morphology 
(globules \#1, \#4 and \#6).
 This is indicative of little active interaction with ionizing EUV photons 
that would have carved them into \mbox{``tear drops''} and less spherical shapes  \citep[e.g.,][]{DeMarco06}. This  would also explain why these 
 globules survive (longer)  in this harsh  environment. Alternatively, some of the more
roundish globules might have already lost their lower density molecular tails \mbox{\citep[e.g.,][]{Gahm13}}. We find this scenario less likely for globules currently embedded in the  shell, but this may be the case for globule \#3 which looks more isolated in its \mbox{position--velocity} diagram (Fig~\ref{fig:pv_3}) and is perhaps located outside the shell. However,  the shell geometry is more complicated than the simple modeled arcs shown in the diagrams, and therefore this argument is not conclusive.
In addition, and even if the detected globules in Orion are all mostly illuminated laterally, projection effects may influence their apparent morphologies 
(that depend on the aspect angle). In extreme cases, an elongated globule with head or tail may look spherical.

Eventually, the ionization front associated with the expanding \HII~region inside the expanding bubble  will encounter these globules, further sculpting, photoevaporating, and compressing them \citep[see simulations in][]{Henney09}. This radiative interaction will lead to  a significant erosion of the  globules \citep[e.g.,][]{Bertoldi90, Lefloch94}. 
The smallest and roundish  globules may develop ``cometary'' or \mbox{``tear drop''} shapes. Still, their morphological evolution depends on the strength and orientation of the magnetic field  threading them. For strong fields, simulations predict that photoevaporating globules can acquire flattened shapes \citep{Henney09}. Only globule \#9 is suggestive of this geometry.

On the other hand, the larger and more elongated CO structures at positive-v$_{\rm LSR}$ (globules \#5, \#8, and \#10) represent the first stages in the formation of (Orion-equivalent) miniature \mbox{``Pillars of Creation''} \citep[e.g.,][]{Pound98,Tremblin12}. Globules \#5 and \#8 are located
at the limb-brightened edge of the shell (and their CO emission 
is connected to that of the background molecular cloud) whereas \#10 is not at the limb and it is more isolated in velocity with respect to the OMC.
These structures, and also the ``KH ripples'' \citep{Berne10}, display bright  8\,$\upmu$m and [\CII]\,158\,$\upmu$m emission  rims pointing toward the Trapezium (see Figs. \ref{fig:globule-images} and  \ref{fig:globule-images-Cplus}). They are also characterized by higher CO excitation temperatures \mbox{($\sim$40-60\,K)} and higher \textit{local} $G_0$ values (up to hundreds). Hence, they are in a more active phase of radiative shaping and new small globules may detach from them.

\subsection{Low-mass star-formation in the Veil?}

    An intriguing question is whether this kind of small globule can form new stars of very low mass. For several of the detected globules in Orion, their external pressure seems sufficient to overcome the stability limit for \mbox{Bonnor-Ebert} spheres
    (see \mbox{Table~\ref{table:globule-masses}})  and so they do not seem to be in hydrostatic equilibrium but  instead to be dynamically evolving \mbox{\citep[e.g.,][]{Galli02}}. 
     However, the estimated globule masses are  smaller than the Jeans mass needed for gravity to dominate and trigger collapse (Sect.~\ref{sec:masses}). Indeed, the inferred  large globule virial parameters, \mbox{$\alpha$\,=\,$5 \sigma^2 R_{\rm g}$\,/\,$G M_{\rm g}$\,$\gg$\,1}, the derived scaling of $\alpha$ with their mass, 
\mbox{$\alpha$\,$\propto$\,$M_{\rm g}^{-0.6\pm0.1}$}, and the likely fact that $P_{\rm ext}$\,$\approx$\,$P_{\rm B,\,g}$ all agree with theoretical expectations of  pressure-confined, gravitationally unbound globules \citep[e.g.,][]{Bertoldi92,Lada08}.

Interestingly, while the dense prestellar cores in Orion's integral-shape filament are  confined by pressure due to the weight of the molecular cloud and the filament \citep{Kirk17}, the CO globules in the Veil are mainly confined by the turbulent pressure of the wind-driven shell. This points to transient molecular globules that would ultimately be photoevaporated \mbox{\citep[e.g.,][]{Oort55,Gorti02}}  or expand and disperse as the pressure  inside the expanding shell decreases.  In this likely scenario, most CO globules, and the bulk material in the shell, will not form new stars.  Given the large masses swept up in the shell, this will limit the global star-formation rate in Orion. 
    
The exception that confirms the rule is globule \#1 (shown in detail in Fig.~\ref{fig:globule-images}). Its position coincides with a known young stellar object that was classified as a pre-main sequence star with a disk and was detected in previous infrared imaging surveys of Orion \citep[\mbox{YSO \#1728} of][]{Megeath12}.  This match suggests that some low-mass protostars do exist inside the shell. In these cases, their molecular cocoons were probably dense and massive enough before being engulfed by the expanding shell. 

The more isolated and smaller CO globules  detected in the shell are part of an interesting class of tiny molecular clouds \mbox{($\sim$0.1\,$M_{\odot}$\,$\simeq$\,100\,$M_{\rm Jup}$)} that are different from the bigger and more massive \mbox{``Bok globules''}. The latter ones usually form one or a few low-mass stars but they are mostly unrelated to \HII~regions \citep[][]{Bok47,Reipurth83,Nelson99,Launhardt10}.
Previous studies of small globules around the \HII~region in the Rosette Nebula have suggested 
that they could be a source of brown dwarfs and free-floating planetary-mass objects \mbox{\citep[e.g.,][]{Gahm07,Gahm13}}. However, our virial 
and Jeans mass analysis does not support this scenario
for the small globules in Orion. Nevertheless, it would be interesting to carry out follow-up deep observations of these objects
in higher critical-density molecular tracers  able to reveal the presence of denser and
more massive gas cores in their interiors.

\section{Summary and conclusions}

We expand previous  maps of Orion~A taken
with the IRAM\,30m telescope in the $^{12}$CO, $^{13}$CO, and C$^{18}$O~($J$\,=\,2-1) lines at 11$''$ resolution ($\simeq$\,4,500\,AU). We investigate a  2\,pc\,$\times$\,4\,pc ($\sim$600\,arcmin$^2$) region  of the neutral shell that confines the  \mbox{wind-driven} Veil bubble around the Trapezium cluster. This massive shell, which is made of swept up material from
the natal Orion molecular cloud, is very bright in [\CII]\,158\,$\upmu$m, PAHs, and 70\,$\upmu$m dust emission
\citep{Pabst19}. 
Owing to intense UV radiation from  \mbox{$\theta^1$ Ori C}, the most massive O-type star in the cluster, the expected column densities of molecular gas in this  extended but translucent foreground material were very low. 
We summarize the primary results of this work as follows:

-- We find that widespread and extended  CO emission is largely absent from the  Veil. This implies that most of the  neutral material  that surrounds  the Orion cluster   is ``CO-dark'' but not necessarily 100\,\% atomic. In particular,  
we present the detection of CO globules (some of them at 
\mbox{negative v$_{\rm LSR}$})
embedded in the expanding shell that encloses the bubble. 

-- The CO globules are small (\mbox{$R_{\rm g}$\,$\simeq$\,7,100\,AU}), moderately dense \mbox{($n_{\rm H}$\,=\,3.6$\cdot$10$^4$\,cm$^{-3}$}), and have low masses:
\mbox{$M_{\rm g}$\,=\,0.3\,$M_{\odot}$} (median values of the sample).
The observed [\CII]\,158\,$\upmu$m, $^{12}$CO, and \mbox{$^{13}$CO (2-1)}  emission  from the \mbox{negative-v$_{\rm LSR}$ globules} (\#1, \#2, and \#3) and also from \#6, those closer to us, are reproduced by depth-dependent PDR models with  $n_{\rm H}$\,$\approx$\,2$\cdot$10$^4$\,cm$^{-3}$  and 
$G_0$\,$\approx$\,40 (both quantities within a factor 3).  These globules show
roundish morphologies.
In the other extreme of the sample, structures \#5 and \#10 (at \mbox{positive v$_{\rm LSR}$}) are bigger and more elongated. Fitting their emission lines 
requires stronger FUV photon fluxes ($G_0$\,$\gtrsim$\,500), higher densities ($n_{\rm H}$\,$\gtrsim$\,4$\cdot$10$^4$\,cm$^{-3}$), and higher extinction depths 
\mbox{($A_{\rm V}$\,$\gtrsim\,$5\,mag)}. These  values are consistent with the
detection of  \mbox{C$^{18}$O (2-1)} emission  from these two molecular gas structures.

-- The inferred \mbox{nonthermal} velocity dispersions 
(\mbox{$\sigma_{\rm turb}$(C$^+$)\,$\simeq$\,1.3\,km\,s$^{-1}$} and
\mbox{$\sigma_{\rm turb}$($^{13}$CO)\,$\simeq$\,0.5\,km\,s$^{-1}$}) 
indicate that the gas in the shell  and in the envelopes around the globules is more turbulent and warmer \mbox{($T_{\rm k}$\,$\approx$100\,K)} than in their molecular  interiors \mbox{($T_{\rm k}$\,$\approx$10--50\,K)}.
These dispersions imply transonic to supersonic motions. The ratio of the \mbox{nonthermal} to thermal pressure is always greater than one (and up to $\sim$10) both in the globule interiors and in their envelopes. We derive that the external  pressure around the CO globules (turbulent\,+\,thermal) is high, $P_{\rm ext}/k$\,$\gtrsim$\,10$^7$\,K\,cm$^{-3}$ (median value). 

-- The magnetic field strength previously derived in \mbox{low-density} portions of the Veil and in the denser gas of the Orion Bar suggests that the CO globules
in the shell  can be threaded by
fields of the order of \mbox{100-200\,$\upmu$G}. These values  would be sufficient to magnetically support them \mbox{($P_{\rm ext}$\,$\simeq$\,$P_{\rm B,\,g})$}. \mbox{Together} with the inferred large virial parameters, $\alpha$\,$\gg$1, the gathered evidence suggests pressure-confined, gravitationally unbound globules.

-- The CO  globules are embedded in the predominantly  neutral material of the shell. They are either transient objects formed by hydrodynamic instabilities or pre-existing over-dense structures of the original molecular cloud.
They are being sculpted by the passing shock associated with the expanding shell and by UV radiation from \mbox{$\theta^1$ Ori C}. Several
of them are isolated and intriguingly roundish. This suggests that they are not being actively photo-ionized.
The more elongated and warmer globules \mbox{(\#5, \#8, and \#10)}  likely represent the first stages in the formation of  (Orion-equivalent) miniature Pillars of Creation.

-- The estimated masses of all globules are lower than their  Jeans masses, and thus  do not imply they will easily form stars unless they  accrete more mass. To break the rule,   \mbox{globule \#1} coincides with the position of a known YSO. This suggests that some \mbox{low-mass} protostars do exist inside the expanding shell. One possibility is that their molecular cores were massive enough and had collapsed before being caught by the shell.
\\

The lack of extended CO emission from the swept-up shell that encloses the  Veil bubble in Orion implies that  winds and UV radiation from young massive stars expel, agitate, and \mbox{photo-dissociate} most of the disrupted molecular cloud gas. This material is blown to large distances, far from the main star-forming cores. Large-scale turbulence and FUV heating across the shell hinders the formation of new stars. Given the large amount of disrupted material,
these  feedback processes must limit the star-formation rate in Orion. M42 is certainly impressive (mostly because it is so close to us) but it is interesting to see that most of the fireworks, energetics, and gas dynamics are mostly  driven by the massive stellar system \mbox{$\theta^1$ Ori C}.\\

\begin{acknowledgements} We acknowledge helpful comments and suggestions from our referee. We warmly thank the operators, AoDs, and chefs at the
IRAM\,30m telescope for
their  support while the CO observations were conducted. This work is also
based on observations made with the NASA/DLR Stratospheric Observatory for Infrared Astronomy (SOFIA). SOFIA is jointly operated by the Universities Space Research Association, Inc. (USRA), under NASA contract \mbox{NNA17BF53C}, and the Deutsches SOFIA Institut (DSI) under DLR contract \mbox{50 OK 0901} to the University of Stuttgart.
 We acknowledge the work, during the C$^+$ upGREAT square degree survey of Orion, of the USRA and NASA staff of the Armstrong Flight Research Center in Palmdale, the Ames Research Center in Mountain View (California), and the Deutsches SOFIA Institut.  We thank the Spanish MICIU for funding support under grant AYA2017-85111-P.  Research on the ISM  at Leiden Observatory is supported through a Spinoza award.
\end{acknowledgements}

\bibliographystyle{aa}
\bibliography{references}

\begin{appendix}\label{Sect:Appendix}

\section{Complementary Tables and Figures}\label{app:tables}

\begin{table*}
\caption{Offsets of the CO globules with respect to $\theta^1$ Ori C  and line peak temperature (absolute calibration error of $\sim$15\%).\label{table:line-peaks}} 
\centering
\begin{tabular}{rcccc@{\vrule height 8pt depth 5pt width 0pt}}
\hline\hline
        &    Offsets             &                  &  $T_{\rm mb,\,P}$ (K)  &                 \\
Globule &                                &  $^{12}$CO (2-1) &  $^{13}$CO (2-1)  & C$^{18}$O (2-1) \\
\hline
 \#1  &   (-825$''$,-122$''$)    &   3.1            &  0.3              &  --            \\   
 \#2  &   (-878$''$,-155$''$)    &   5.3            &  --               &  --             \\
 \#3  &   (-999$''$,-367$''$)    &   8.8            &  0.8              &  --            \\
 \#4  &   (-1143$''$,-1249$''$)  &  14.6            &  0.8              &  --            \\
 \#5  &   (-1762$''$,-654$''$)   &  40.7            & 13.3              & 1.4           \\
 \#6  &   (-1108$''$,-296$''$)   &   3.4            &  0.3              &  --             \\
 \#7  &   (-992$''$,-202$''$)    &  40.8            &  8.9              &  --            \\
 \#8  &   (-1105$''$,-54$''$)    &  33.2            &  5.7              &  --          \\ 
 \#9  &   (-1023$''$,-35$''$)    &  26.3            &  2.2              &  --            \\
\#10  &   (-841$''$,+61$''$)     &  52.5            &  15.5             &  1.4          \\  
\hline
\end{tabular}
\tablefoot{$^{12}$CO, $^{13}$CO, and C$^{18}$O ($J$=2-1) lines convolved at a uniform angular and spectral resolution of 16$''$ and 0.4\,km\,s$^{-1}$, respectively.}
\end{table*}

\begin{table*}
\caption{Line  parameters of the CO globules obtained from Gaussian fits to the spectral component blueshifted from OMC-1 velocities.\label{table:line-fits}} 
\centering
\begin{tabular}{rcccccccccc@{\vrule height 8pt depth 5pt width 0pt}}
\hline\hline
      &    Vel. Centroid    &            &            & Line width          &             &           & Line intensity      &            &           \\
      &    (km\,s$^{-1}$)   &            &            & (km\,s$^{-1}$)      &             &           & (K\,km\,s$^{-1}$)   &            &           \\
globule & [\CII]                            & $^{12}$CO  & $^{13}$CO  & [\CII]              & $^{12}$CO   & $^{13}$CO & [\CII]              & $^{12}$CO  & $^{13}$CO \\
\hline
 \#1  & -2.1(0.3)           & -2.4(0.5)  & -2.3(0.5)  & 5.4(0.6)            & 1.4(0.1)    & 1.3(0.5)  & 32.4(2.8)           &  4.8(3.2)  & 0.5(0.2)\\   
 \#2  & -1.2(0.2)           & -2.0(0.3)  & --         & 6.6(0.5)            & 1.9(0.7)    & --        & 33.1(2.0)           & 10.6(3.3)  & --       \\
 \#3  & -5.2(0.1)           & -5.4(0.1)  & -5.4(0.1)  & 2.5(0.3)            & 1.3(0.1)    & 1.1(0.1)  & 22.4(2.1)           & 13.2(0.2)  &  0.9(0.1)\\
 \#4  & +3.4(0.1)           & +3.7(0.1)  & +3.5(0.1)  & 2.4(0.2)            & 1.6(0.1)    & 1.2(0.1)  & 58.4(3.0)           & 24.3(1.3)  & 1.0(0.1)\\
 \#5  & +6.0(0.1)           & +6.0(0.1)  & +6.4(0.1)  & 6.0(0.2)            & 3.1(0.1)    & 2.0(0.1)  & 112.0(2.9)          & 135.3(0.2) & 28.7(0.1)\\
 \#6  & +1.6(0.1)           & +0.8(0.4)  & +1.9(0.1)  & 2.8(0.2)            & 1.1(0.3)    & 1.2(0.6)  & 47.7(2.3)           & 3.9(1.3)   & 0.4(0.1)\\
 \#7  & +5.4(0.1)           & +4.7(0.3)  & +4.7(0.1)  & 3.5(0.2)            & 1.5(0.3)    & 1.1(0.1)  & 80.8(5.7)           & 64.1(2.6)  & 10.2(0.1)\\
 \#8  & +3.7(0.3)           & +3.8(0.3)  & +4.0(0.1)  & 3.0(0.4)            & 1.9(0.3)    & 1.0(0.1)  & 29.2(7.3)           & 65.4(3.2)  & 6.1(0.2)\\ 
 \#9  & +2.7(0.2)           & +3.1(0.1)  & +3.1(0.1)  & 2.5(0.3)            & 1.5(0.1)    & 1.4(0.1)  & 18.7(4.3)           & 41.5(0.2)  & 3.2(0.2)\\
\#10  & 2.8(0.1)            & +2.7(0.1)  & +2.8(0.1)  & 4.9(0.1)            & 2.0(0.1)    & 1.3(0.1)  & 140.3(2.3)          & 111.9(0.2) & 21.2(0.2)\\  
\hline
\end{tabular}
\tablefoot{[\CII]\,158\,$\upmu$m,  $^{12}$CO, and $^{13}$CO ($J$=2-1) lines convolved at a uniform angular and spectral resolution of 16$''$ and 0.4\,km\,s$^{-1}$, respectively.}
\end{table*}

\begin{figure*}[h]
\vspace{2cm} 
\centering   
\includegraphics[scale=0.207, angle=0]{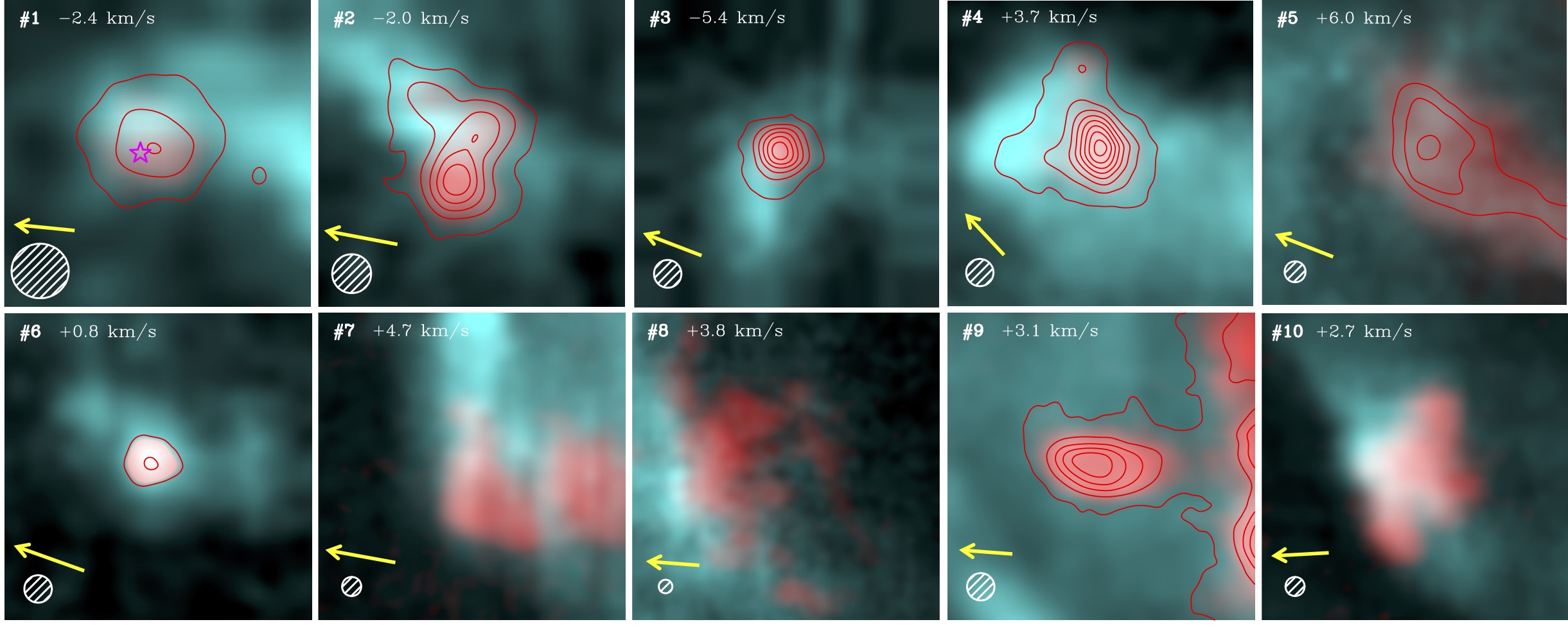}
\caption{Gallery of detected CO globules. Same as Fig.~\ref{fig:globule-images} but showing the [\CII]\,158\,$\upmu$m emission (bluish) integrated in exactly the same velocity range
as the CO emission from each globule (reddish). The yellow arrows indicate the direction to star \mbox{$\theta^1$ Ori C} in the Trapezium.}\label{fig:globule-images-Cplus} 
\end{figure*}

\begin{figure}[ht]
\centering   
\includegraphics[scale=0.26, angle=0]{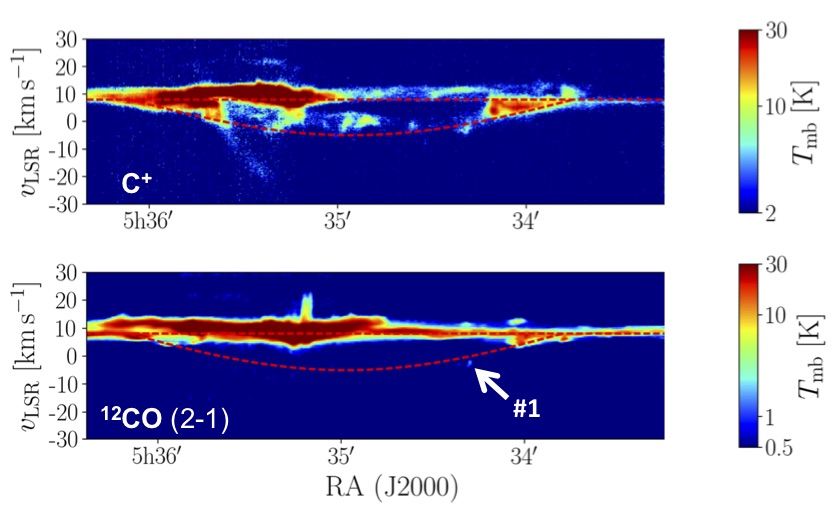}
\caption{Same as Fig.~\ref{fig:pv_4} but for an east--west cut across the bubble at the declination of globule \#1 (-$5^o25'24.23''$).}\label{fig:pv_1}
\end{figure}

\begin{figure}[ht]
\centering   
\includegraphics[scale=0.26, angle=0]{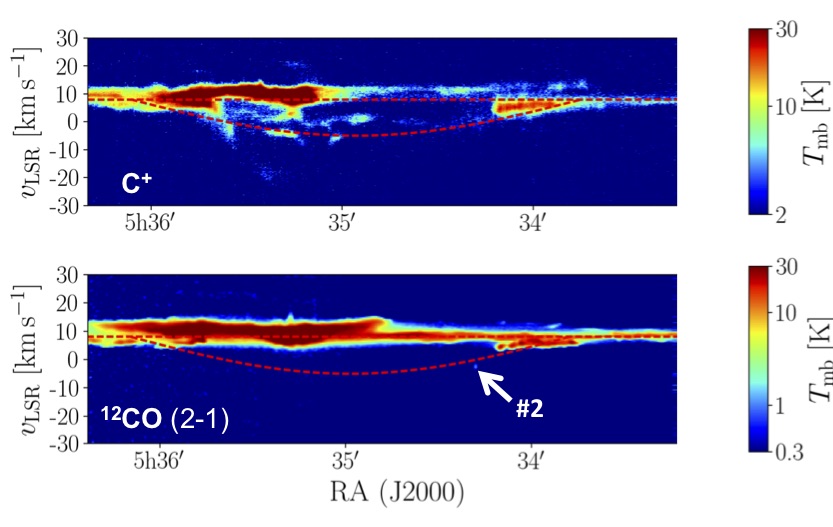}
\caption{Same as Fig.~\ref{fig:pv_4} but for an east--west cut across the bubble at the declination of globule \#2 (-$5^o25'57.52''$).}\label{fig:pv_2}
\end{figure}

\begin{figure}[ht]
\centering   
\includegraphics[scale=0.26, angle=0]{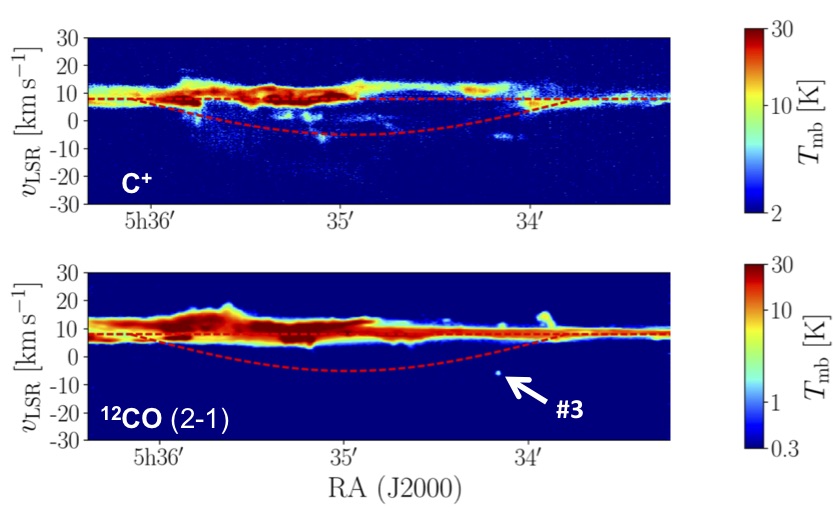}
\caption{Same as Fig.~\ref{fig:pv_4} but for
an east--west cut across the bubble at the declination of globule \#3 (-$5^o29'29.40''$).}\label{fig:pv_3}
\end{figure}

\begin{figure}[ht]
\centering   
\includegraphics[scale=0.26, angle=0]{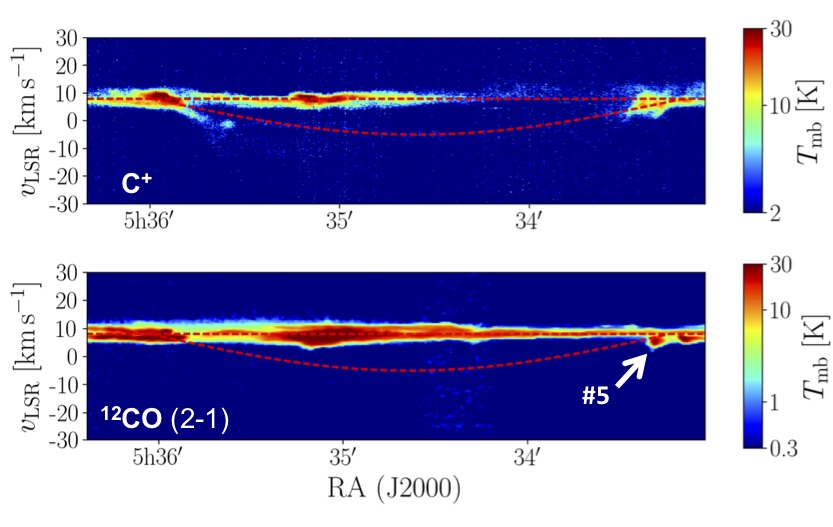}
\caption{Same as Fig.~\ref{fig:pv_4} but for
an east--west cut across the bubble at the declination of globule \#5 (-$5^o34'16.29''$).}\label{fig:pv_5}
\end{figure}

\begin{figure}[ht]
\centering   
\includegraphics[scale=0.26, angle=0]{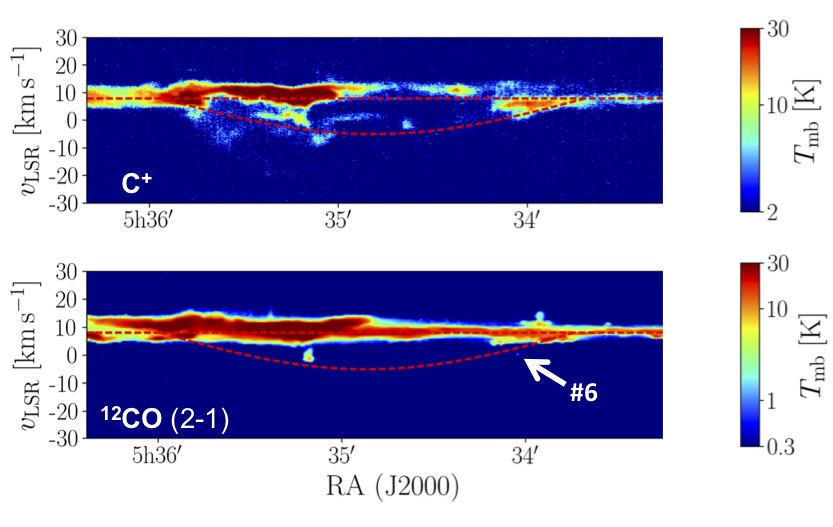}
\caption{Same as Fig.~\ref{fig:pv_4} but for
an east--west cut across the bubble at the declination of globule \#6 (-$5^o28'18.58''$).}\label{fig:pv_6}
\end{figure}

\begin{figure}[ht]
\centering   
\includegraphics[scale=0.26, angle=0]{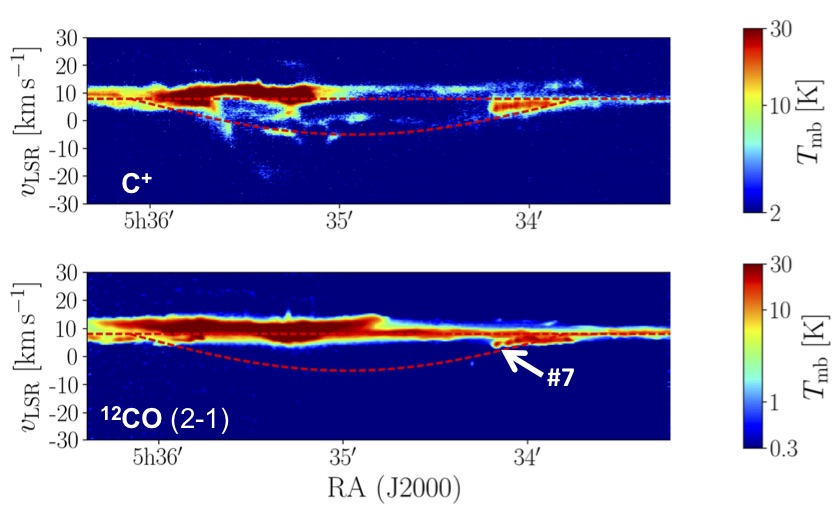}
\caption{Same as Fig.~\ref{fig:pv_4} but for
an east--west cut across the bubble at the declination of globule \#7 (-$5^o26'44.24''$).}\label{fig:pv_7}
\end{figure}

\begin{figure}[ht]
\centering   
\includegraphics[scale=0.26, angle=0]{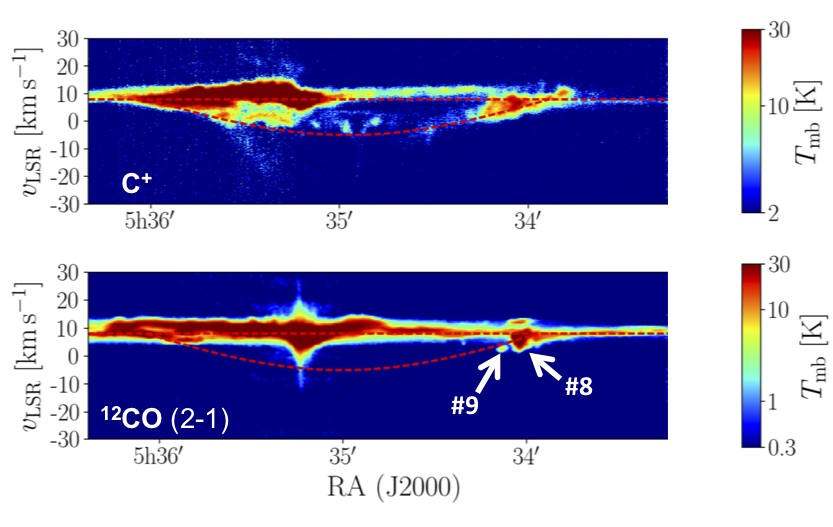}
\caption{Same as Fig.~\ref{fig:pv_4} but for
an east--west cut across the bubble at the declination of globules \#8 and \#9 (-$5^o24'$).}\label{fig:pv_89}
\end{figure}

\begin{figure}[ht]
\centering   
\includegraphics[scale=0.26, angle=0]{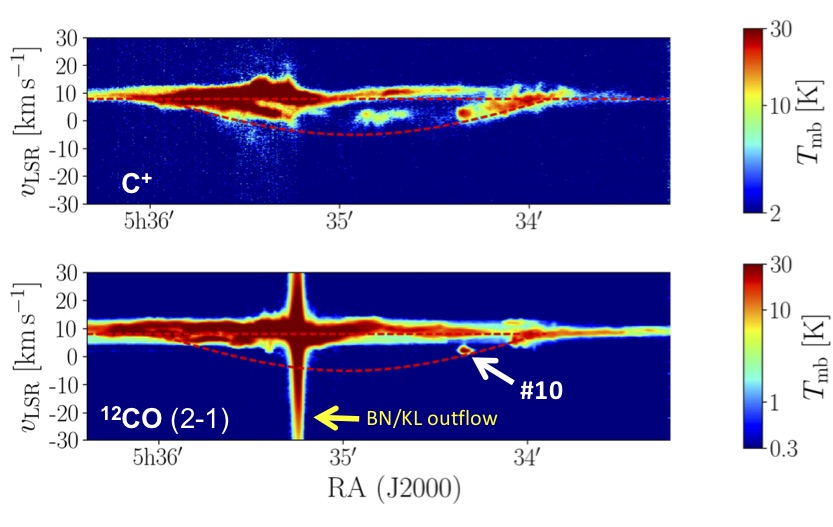}
\caption{Same as Fig.~\ref{fig:pv_4} but for
an east--west cut across the bubble at the declination of globule \#10 (-$5^o22'21.93''$).}\label{fig:pv_10}
\end{figure}

\begin{figure*}[]
\centering   
\includegraphics[scale=0.50, angle=0]{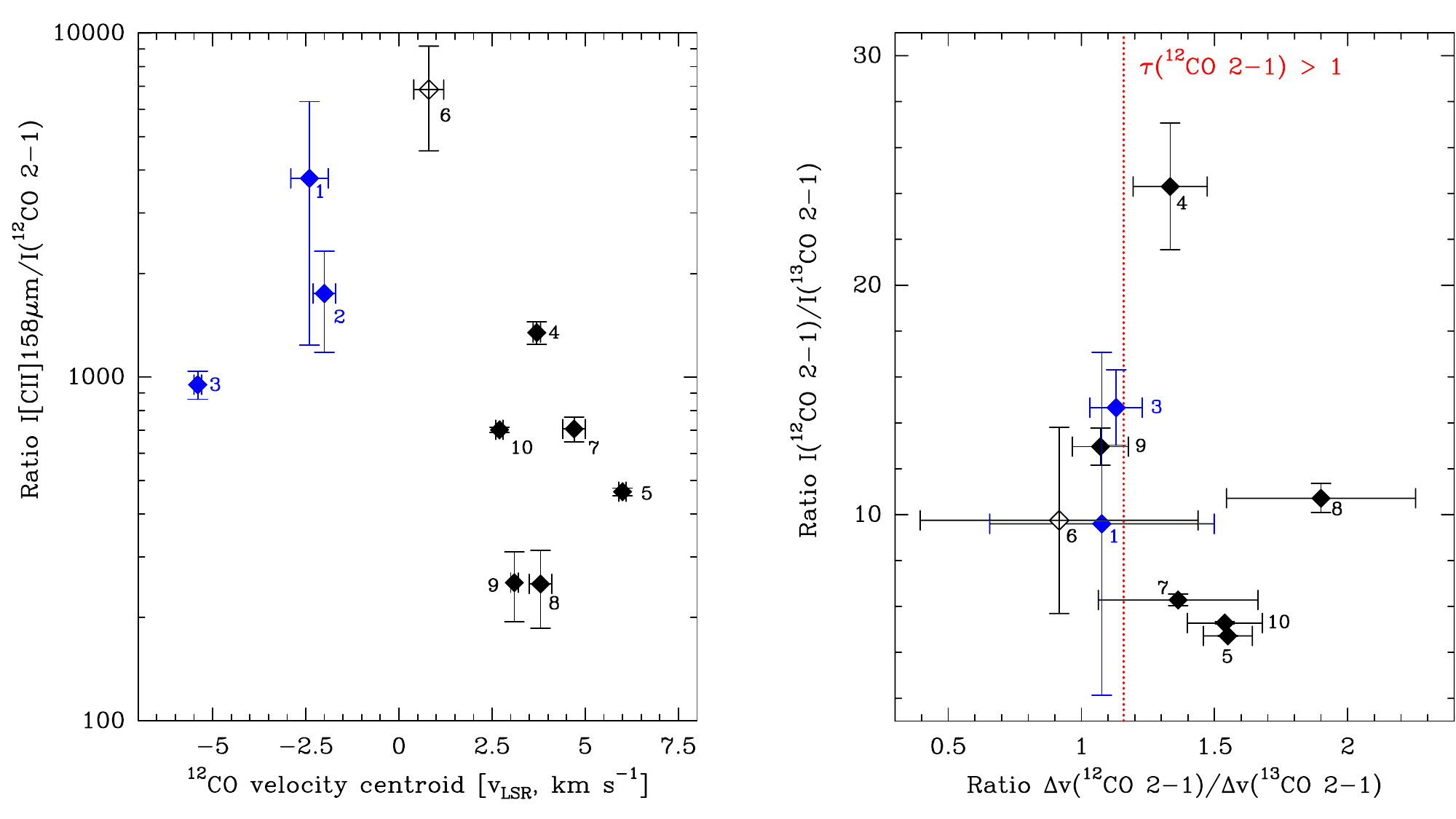}
\caption{CO line emission from the globules. Blue (black) points correspond to the negative (positive) LSR velocity globules. Globule \#6 lies close in velocity to the negative-v$_{\rm LSR}$ globules and indeed shows similar emission properties. \textit{Left panel:} \mbox{[\CII]158\,$\upmu$m/$^{12}$CO (2-1)}
line intensity ratio, in units of erg\,s$^{-1}$\,cm$^{-2}$\,sr$^{-1}$, as a function of globule velocity centroid (as observed in $^{12}$CO). \textit{Right panel:} $^{12}$CO/$^{13}$CO (2-1) intensity ratio as a function of their line-width ratio. As gas column density raises, opacity broadening increases the width of $^{12}$CO lines. 
 The dashed red line approximately marks the \mbox{$\tau$($^{12}$CO 2-1)$>$1} transition. Globules at the left of this line are not very thick in $^{12}$CO~(2-1), yet they show very low $^{12}$CO/$^{13}$CO\.(2-1) line intensity ratios produced by $^{13}$CO chemical fractionation.}\label{fig:ratios}
\end{figure*}

\begin{figure*}[]
\centering   
\includegraphics[scale=0.40, angle=0]{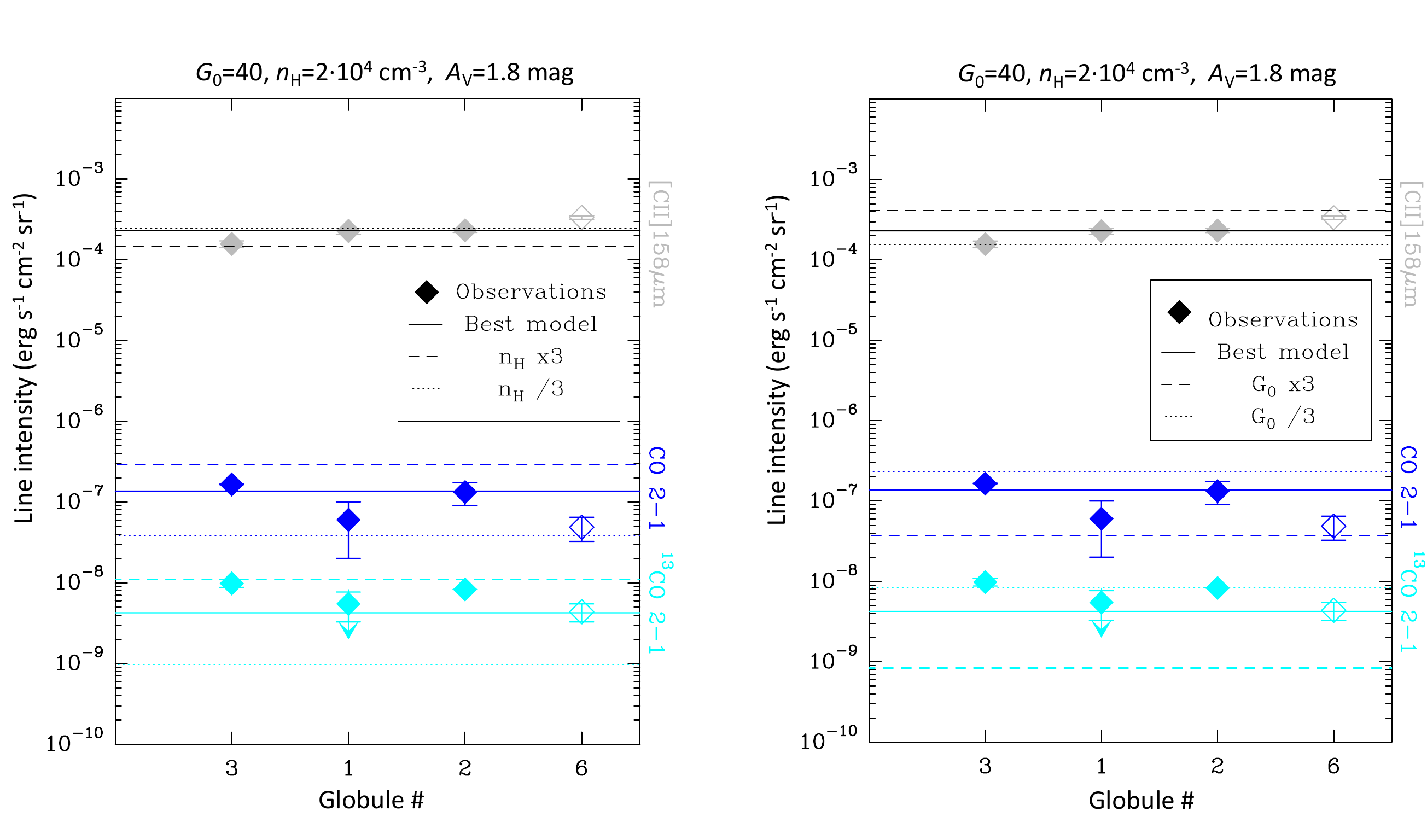}
\caption{Comparison of PDR model predictions and observed line intensities from the \mbox{negative-v$_{\rm LSR}$} globules. Error bars show the 1$\sigma$ uncertainty to the measured intensities obtained from a Gaussian fit to the observed lines. Globule \#6 lies close in LSR velocity and shows similar emission properties to the \mbox{negative-v$_{\rm LSR}$} globules. The velocity centroid of each globule increases from left to right in the x-axis (see Fig.\ref{fig:spectra}). \textit{Left} panel shows variations in the gas density whereas
the  \textit{right} panel shows variations in $G_0$.} \label{fig:pdr_mod_inten}
\end{figure*}

\end{appendix}

\end{document}